\documentclass[a4paper,11pt]{article}
\usepackage[a4paper, total={6in, 10in}]{geometry}
\usepackage[table,xcdraw]{xcolor}
\usepackage{subcaption}
\usepackage{amssymb}
\usepackage{rotating}
\usepackage{amsfonts}
\usepackage[official]{eurosym}
\usepackage{stackrel}
\usepackage{caption}
\usepackage{graphicx}
\usepackage{xspace}
\usepackage{tikz}
\usepackage{mathtools}
\usepackage{url}
\usepackage{authblk}
\usepackage{hyperref}
\usepackage{amsthm}
\usepackage{bm}
\usepackage[title]{appendix}
\usepackage{bbold}
\usepackage{pdfpages}
\usepackage{csquotes}

\theoremstyle{definition}
\newtheorem{hyp}{Hypothesis}

\usepackage[english]{babel}

\usepackage{float}

\usepackage[
    backend=biber,
    style=authoryear,
  ]{biblatex}

\mathtoolsset{showonlyrefs,showmanualtags}

\hypersetup{
	colorlinks=true,
	urlcolor=blue,
	citecolor=blue,
	linktoc=all,
	linkcolor=blue}
	
\definecolor{DarkGreen}{rgb}{0,0.4,0}
\definecolor{cobalt}{rgb}{0.0, 0.28, 0.67}

\title{A constraint on the dynamics of wealth concentration}

\author{Valerio Astuti}

\date{}

\begin{document}

\maketitle

\begin{abstract}
In the context of a large class of stochastic processes used to describe the dynamics of wealth growth, we prove a set of inequalities establishing necessary and sufficient conditions in order to avoid infinite wealth concentration. These inequalities generalize results previously found only in the context of particular models, or with more restrictive sets of hypotheses. In particular, we emphasize the role of the additive component of growth - usually representing labor incomes - in limiting the growth of inequality.
Our main result is a proof that in an economy with random wealth growth, with returns non-negatively correlated with wealth, an average labor income growing at least proportionally to the average wealth is necessary to avoid a runaway concentration.
One of the main advantages of this result with respect to the standard economics literature is the independence from the concept of an equilibrium wealth distribution, which does not always exist in random growth models. 
We analyze in this light three toy models, widely studied in the economics and econophysics literature.
\end{abstract}

\section{Introduction}
The concentration of wealth and income is an increasingly studied problem in economics (\cite{piketty2014inequality, de2015thomas, gabaix2016dynamics, brandolini2019distribution}). One of the causes for the growing attention is that, after a sharp decrease starting in the first half of the twentieth century, wealth concentration is growing again at a worrying pace at least from the 1980's.
An important part of the research work on the subject is directed toward the study of the mechanisms and causes of such a phenomenon, and possibly on how to limit the increase of inequality.  
One of the approaches to describe the properties of wealth distribution is to introduce models of economic agents, and simulate (or solve, when possible) the dynamics of these models to study the asymptotic equilibrium states. Commonly studied models are characterized by different assumptions, level of details, and methods of solution, often arriving also at different conclusions. 

A large branch of research in the economic literature focuses on the microfoundations of wealth dynamics and a close replication of real data, at the cost of detailed hypotheses on agents' choices and fine calibration processes (\cite{davies2000distribution, cagetti2008wealth, de2015quantitative, gabaix2016dynamics, xavier2021wealth}). A common feature of this type of models is the characterization of the consumption choices of agents by a postulated form of utility function: the agents decide how much of their wealth to consume with the maximization of some function representing their preferences (\cite{modigliani1954utility}). 
Picking a specific form of utility function amounts to a detailed specification of the consumption preferences for all the agents in the model, and as such it represents a very strong assumption on the microfoundations of wealth dynamics. 
One of the recent findings of this line of research is the role in the high wealth concentration of the so-called \emph{type} and \emph{scale dependence} (\cite{gabaix2016dynamics, xavier2021wealth}). 
Considering a positive correlation between returns at different times (type dependence) and between returns and net wealth (scale dependence), it is possible to explain in a more realistic way the behaviour of the wealth distribution for high values of wealth. 
A necessary hypothesis to perform this type of analysis is the existence of an asymptotic equilibrium distribution of wealth. After a long enough period of time the dynamics of the model converges to this equilibrium distribution, and properties like inequality can be studied for this distribution. 

A second kind of models, closer to the physics literature and modeling style, emphasizes frugality of hypotheses and ease of interpretation (\cite{bouchaud2000wealth, cardoso2020wealth, gere2021wealth}). 
An approach in this line of research, not based on a detailed description of agents' choices but on general properties of the evolution of wealth over time - as, for example, the multiplicative stochastic nature of the growth process - has been used to prove some interesting results (\cite{biondi2020financial, gere2021wealth, cardoso2021wealth}). 
In these models the main driver of growth in inequality is often the random heterogeneity in returns, and it is enough to drive the concentration of wealth to arbitrarily high levels. In particular it is a well-established result that for \emph{i.i.d.} returns on wealth and labor incomes, in order to have a non-trivial equilibrium wealth distribution the dynamics has to be contracting on average, and the average labor income must have a fixed ratio with the average wealth (\cite{kesten1973random}). 
A similar behaviour is noted, in the context of specific models, also in \cite{bouchaud2000wealth, gabaix2016dynamics, cardoso2020wealth, biondi2020financial}. More generally the role of uncertainty in wealth concentration is pointed out in \cite{levy2003investment, levy2003rich}. 
This branch of research does not usually assume the existence of an equilibrium, neither in terms of an asymptotic wealth distribution, nor in terms of wealth inequality. 
In many of these models an infinite concentration of wealth is found in the long time limit of the evolution. 
Results from this second branch of literature, however, often do not take into account structural differences between agents. 
It is thus possible to ask how the more detailed features contemplated in the economic models can influence the general, model-independent results derived without emphasis on a particular microfounding structure.
In particular, it can be interesting to study the features separating the long-time evolution of these two classes of models, and identify the factors constraining inequality in the various models.  
In this paper we study processes composed of a multiplicative term representing financial returns and an additive term representing labor income. 
In the context of this class of models we characterize the necessary and sufficient conditions to have a bounded wealth inequality, and find these conditions to be related to the ratio between the additive (labor income) term and the multiplicative (financial wealth) term. 
These conditions are valid for models closer to the economic literature, in which consumption is derived from a given form of utility function, but are more general, in that we take into consideration only aggregate consumption functions, which are not necessarily derived from a maximization process.
In addition, the derivation of these conditions is not dependent on the existence of an asymptotic wealth distribution, thus it is valid also in contexts in which the traditional methods of analysis fail to apply.

Our results put a lower bound on the ratio between average labor income and average wealth. 
To avoid infinite concentration of wealth, this ratio must remain constant over time, which implies that the average labor income has to grow at least proportionally to the average wealth. 
The finding that without a sufficiently growing labor income the concentration of wealth is bound to grow is reminiscent of the main result of \cite{piketty2014inequality, piketty2015wealth}. There, from a study of the historical evolution of wealth inequality, the authors conclude that a condition to have non-increasing inequality is labor productivity growing faster than the average financial returns. 
Interpreting this result in our terms, we show that a channel through which productivity growth can hinder wealth concentration is the increase of the average labor income. 

\section{Hypotheses of the derivation}
\label{sec:hypotheses}
Our results are valid for a large class of models of stochastic wealth growth, which describe the evolution of wealth in a population of agents in the form:\footnote{The symbol $\hat{\bullet}$ is used to denote stochastic variables.}
\begin{equation}
\label{general_stochastic_growth}
    \hat{x}_{t+1}^i = \hat{A}_t^i(x_t, \xi_t)\,x_t^i + \hat{B}_t^i(x_t, \xi_t)
\end{equation}
where the index $i$ denotes the agent whose wealth we are evaluating, and $\hat{A}_t^i(x_t, \xi_t)$ and $\hat{B}_t^i(x_t, \xi_t)$ are stochastic processes that can in general depend on the wealth value $x_t$ and a set of additional parameters $\xi_t$. 
These parameters can represent, for example, consumption choices of each agent or heterogeneity in investment choices. 
We assume the processes $\hat{A}_t(x_t, \xi_t)$ and $\hat{B}_t(x_t, \xi_t)$ and $\hat{x}_{t+1}$ to be non-negative random variables.
In addition, we assume all the random variables considered to have a smooth probability density.

One of the most studied examples of such models describes the evolution of wealth with stochastic processes for returns on wealth and labor incomes, and individual choices of the agents for the amount of wealth to consume:
\begin{equation}
\label{stochastic_growth_1}
    \hat{x}_{t+1}^i = e^{\hat{r}_t^i}\, \left( x_t^i - c_t^i \right) + \hat{y}_t^i
\end{equation}
where $\hat{r}_t^i$ are stochastic returns on wealth, $\hat{y}_t^i$ stochastic labor incomes (usually a process with memory: future values of income are influenced by past values), and $c_t^i\leq x^i_t$ are individual consumption choices. 
If labor incomes have a memory, their realization at time $t$ depends on past realizations (and possibly other variable). In general we write this dependence as:
\begin{equation}
    \hat{y}_t^i = \hat{y}_t^i\left(x_t, \xi_t \right)
\end{equation}
The most studied forms of consumption choices in the economics literature are derived from the maximization of some utility function for the agents (\cite{gabaix2016dynamics, carroll2004theoretical, benhabib2015wealth}). This introduces a dependence on some parameters, describing for example the expectations of each agent about its future wealth. With this in mind, we write consumption as:
\begin{equation}
    c_t^i = c_t^i\left(x_t, \xi_t \right)
\end{equation}
Writing the consumption term as relative to the available wealth, and defining the saving function:
\begin{equation}
    s_t^i(x_t, \xi_t) \coloneqq  \left(1 - \frac{c_t^i\left(x_t, \xi_t \right)}{x_t^i} \right) 
\end{equation}
we can write process \eqref{stochastic_growth_1} as:
\begin{equation}
    \hat{x}_{t+1}^i = e^{\hat{r}_t^i}\, s_t^i(x_t, \xi_t)\,x_t^i + \hat{y}_t^i\left(x_t, \xi_t \right)
\end{equation}
In section \ref{sec:applications} we will focus on models of the form \eqref{stochastic_growth_1}, but for now we only note that they are a partial specification of the general expression \eqref{general_stochastic_growth}, and describe our hypotheses for this more general kind of process. 

A population of agents, each evolving with a stochastic law of the form \eqref{stochastic_growth_1}, implies we have to describe different kinds of probability distributions. At each time $t$ and for each agent $i$, we have the probability distribution related to the realizations of the (possibly dependent) stochastic processes $\hat{A}_t^i(x_t, \xi_t)$ and $\hat{B}_t^i(x_t, \xi_t)$. 
At the same time we want to consider the distribution of wealth $x_t$ and parameters $\xi_t$ in the population of agents (though not strictly a probability distribution, we can always consider the random extraction of an agent, and define the probability for it to have wealth $x_t^i = x$ and parameters $\xi_t^i=\xi$). 
We denote this distribution as:
\begin{equation}
q_t\left(x, \xi \right) = \frac{d^2\, \textbf{P}\left(x_t^i \leq x \, , \, \xi_t^i \leq \xi \right)}{d x \, d \xi} \, 
\end{equation}
Where $\textbf{P}\left(x_t^i \leq x \, , \, \xi_t^i \leq \xi \right)$ is the probability to have wealth $x_t^i \leq x$ and parameters $\xi_t^i \leq \xi$ in the population.
From $q_t\left(x, \xi \right)$ we can derive the distribution of wealth in our population by marginalizing over the additional parameters:
\begin{equation}
    p_t(x) \coloneqq \int d\xi \, q_t\left(x, \xi \right) 
\end{equation}
 In addition we will use the distribution of parameters conditional on the value of wealth:
 \begin{equation}
     h_t(\xi | x) = \frac{q_t\left(x, \xi \right)}{p_t(x)}
 \end{equation}
Finally, we can combine the distributions of the stochastic processes with the ones in the population of agents to obtain, for example, the expectation value of $\hat{A}_t(x_t, \xi_t)$ conditional to a given values of $x_t$. 
In this way we can define the (deterministic) functions of wealth:\footnote{As an example, let's consider for $\hat{B}_t(x_t, \xi_t)$ a labor income process $\hat{y}_t$ independent of wealth and distributed log-normally around its previous realization:
\begin{equation}
    \hat{y}_t = y_{t-1}\,e^{\hat{z}} \quad , \quad \hat{z} \sim N\left(\mu_z , \sigma_z\right)
\end{equation}
In this case the parameters $\xi_t$ are the previous realizations of the stochastic process $\hat{y}_t$.
To obtain the function $\beta_t(x)$ we first have to evaluate the expectation value of $\hat{y}_t$ for a given value of $y_{t-1}$:
\begin{equation}
    \textbf{E}_{\rho_{\hat{y}_t(y_{t-1})}}\left[ \hat{y}_{t}  \right] = y_{t-1}\,e^{\mu_z +  \frac{\sigma_z^2}{2}}
\end{equation}
Second, we evaluate the average of these expectation values in the population of agents having wealth $x_t = x$, with the distribution $h_t(y_{t-1} | x)$:
\begin{equation}
    \beta_t(x) = e^{\mu_z +  \frac{\sigma_z^2}{2}}\, \textbf{E}_{h_t(y_{t-1} | x)}\left[ \left. y_{t-1} \right| x \right]. 
\end{equation}}
\begin{equation}
    \alpha_t(x) \coloneqq \textbf{E}\left[ \left. \hat{A}_t(x_t, \xi_t) \right| x_t=x \right] 
\end{equation}
\begin{equation}
    \beta_t(x) \coloneqq \textbf{E}\left[ \left. \hat{B}_t(x_t, \xi_t) \right| x_t=x \right] 
\end{equation}
\begin{equation}
    \gamma_t(x) \coloneqq \textbf{E}\left[ \left. \hat{x}_{t+1} \right| x_t=x \right] 
\end{equation}
These are the averages over the population of agents having wealth $x_t$ at time $t$, of the expectation values for these agents of the processes $\hat{A}_t(x_t, \xi_t)$, $\hat{B}_t(x_t, \xi_t)$ and $\hat{x}_{t+1}$.
The definitions of $\alpha_t(x)$, $\beta_t(x)$ and $\gamma_t(x)$ imply the identity:
\begin{equation}
    \gamma_t(x) = \alpha_t(x)\, x + \beta_t(x)
\end{equation}
Finally, we denote the probability distribution of having wealth $x_{t+1} = x'$ at time $t+1$, conditional to having wealth $x_t=x$ at time $t$, as $w_t(x'|x)$:
\begin{equation}
    w_t(x'|x) \coloneqq \textbf{E}\left[ \left. \mathbb{1}\left( \hat{x}_{t+1} = x' \right) \right| x_t=x \right] 
\end{equation}
Stated in terms of these functions, the necessary hypotheses to derive our main results are the following:
\begin{hyp}
\label{hyp:positivity_coefficients}
The derivatives of the two functions $\alpha_t(x)$, $\beta_t(x)$ are non-negative functions of wealth:
            \begin{equation}
            \quad \frac{d \alpha_t(x)}{d x}\geq 0 \quad , \quad \frac{d \beta_t(x)}{d x}\geq 0
            \end{equation}
\end{hyp}

\begin{hyp}
\label{hyp:positive_dispersion}
The process $\hat{A}_t(x_t, \xi_t)$ has a positive dispersion for large values of $x_t$:
\begin{equation}
    \exists \; \Gamma_t > 0 \, : \, \lim_{x\to \infty} \textbf{E}\left[ \left. \left| \hat{A}_t(x_t, \xi_t) - \alpha_t(x)\right| \right| x_t=x \right] \geq \Gamma_t.
\end{equation}
\end{hyp}
\noindent
We will point out during the derivation which steps depend on each assumption, and in section \ref{sec:applications} we will show that processes in class \eqref{stochastic_growth_1} comply with all the hypotheses listed here, under weak assumptions on the stochastic processes for returns and labor incomes, and for a wide class of consumption choices.

\section{Evolution of the Gini index}
\label{sec:gini_evolution}
We want to study the evolution of the Gini coefficient in term of the quantities defined in the previous section. The Gini index can be defined as:
\begin{equation}
    \textbf{G}_{x,t} \coloneqq \frac{1}{2\,\textbf{E}_{p_t}\left[ x \right]} \int dx dy\, p_t(x) p_t(y)\, |x-y|
\end{equation}
where we defined the expectation value with respect to the probability distribution $p_t(x)$, introduced in section \ref{sec:hypotheses}:
\begin{equation}
    \textbf{E}_{p_t}\left[ f(x) \right] \coloneqq \int_0^{\infty} dx \, p_t(x) f(x)
\end{equation}
In appendix \ref{app:gini_decomposition}, we prove that in general the evolution of the Gini index can be decomposed as:
\begin{equation}
    \textbf{G}_{x, t+1} = \textbf{G}_{\gamma, t} + \textbf{G}_{\Gamma, t} 
\end{equation}
where $\textbf{G}_{\gamma, t}$ represent the inequality in the conditional average of the evolution of wealth:
\begin{equation}
\label{average_gini}
    \textbf{G}_{\gamma, t} \coloneqq \frac{1}{2\, \textbf{E}_{p_{t}}\left[ \gamma_t(x) \right]   } \int dx dy\, p_{t}(x) p_{t}(y) \, \left|\gamma_t(x) - \gamma_t(y) \right|
\end{equation}
and $\textbf{G}_{\Gamma, t}$ is a term related to the dispersion of the stochastic wealth growth:
\begin{equation}
    \textbf{G}_{\Gamma, t}  \coloneqq \frac{\mathbf{E}_{p_t}\left[F_t\left( x, y\right) \right]}{\textbf{E}_{p_{t}}\left[ \gamma_t(x) \right]  }
\end{equation}
with
\begin{equation}
     \mathbf{E}_{p_t}\left[ F_t(x,y) \right] \coloneqq \frac{1}{2}\int dx dy\, p_{t}(x) p_{t}(y)\, F_t(x,y) 
\end{equation}
\begin{align}
    F_t(x,y) \coloneqq  \, \mathbb{1}\left( x \geq y \right)  & \,\int_{y'>x'} dx'dy'\,w_t(x'|x) w_t(y'|y) \,|x' - y'| \, + \\
    & + \mathbb{1}\left( y > x \right)\, \int_{x'>y'}  dx'dy'\,w_t(x'|x) w_t(y'|y) \,|x' - y'|
\end{align}
While $\mathbf{E}_{p_t}\left[ F_t(x,y) \right]$ is always non-negative, whenever hypothesis \ref{hyp:positive_dispersion} is valid we prove in appendix \ref{app:f_bounds} the existence of a positive lower bound, with the form:
\begin{equation}
\label{eq:main_inequality}
\mathbf{E}_{p_t}\left[ F_t(x,y) \right] \geq \Omega\left( \Gamma_t\right)  \kappa\, \textbf{E}_{p_{t}}\left[ x \right]\, \textbf{P}_{t}\left( x \geq \kappa\, \textbf{E}_{p_{t}}\left[ x \right] \right)^2
\end{equation}
where $\Omega\left( \Gamma_t\right)$ is a positive constant related to the dispersion of the wealth growth process $\Gamma_t$.  
We defined the probability to have wealth greater than a given threshold $\kappa\, \textbf{E}_{p_{t}}\left[ x \right]$ for some $\kappa > 0$ at time $t$:
\begin{equation}
    \textbf{P}_{t}\left( x \geq \kappa\, \textbf{E}_{p_{t}}\left[ x \right] \right) \coloneqq \int_{\kappa\, \textbf{E}_{p_{t}}\left[ x \right]}^{\infty} dx' \, p_t(x')\, x'
\end{equation}
It is worth noticing that the right hand side of inequality \eqref{eq:main_inequality}, however small, is always greater than zero as long as the Gini index is less than one. 
In addition, in appendix \ref{app:f_bounds} we prove the existence of an upper bound for $\mathbf{E}_{p_t}\left[ F_t(x,y) \right]$, with the form:
\begin{equation}
\label{eq:f_upper_bound}
    \mathbf{E}_{p_t}\left[ F_t(x,y) \right] \leq 2 \,\int_0^{\infty}\, dy\, p_t(y)\,P_t\left( x\geq y\right)\, \sigma_t(y) 
\end{equation}
where $\sigma(y)$ is the dispersion of the conditional probability $w_t\left(y' |y \right)$ around its mean:
\begin{equation}
    \sigma_t(y) \coloneqq \int_0^{\infty} dy'\,w_t(y'|y)  \,|y' - \gamma_t(y)|
\end{equation}

We now focus on the form of $\textbf{G}_{\gamma, t}$ for models in the class \eqref{general_stochastic_growth}. 
Hypothesis \ref{hyp:positivity_coefficients} allows us to write equation \eqref{average_gini} as:
\begin{equation}
    \textbf{G}_{\gamma, t} = \left(1 - \frac{\textbf{E}_{p_t}\left[  \beta_t(x) \right] }{\textbf{E}_{p_t}\left[  \gamma_t(x) \right]  } \right) \textbf{G}_{\alpha x, t} + \frac{\textbf{E}_{p_t}\left[  \beta_t(x) \right] }{\textbf{E}_{p_t}\left[  \gamma_t(x) \right]  } \textbf{G}_{\beta, t}
\end{equation} 
with
\begin{equation}
    \textbf{G}_{\alpha x, t} \coloneqq \frac{1}{2\, \textbf{E}_{p_{t}}\left[ \alpha_t(x)\,x \right]   } \int dx dy\, p_{t}(x) p_{t}(y) \, \left| \alpha_t(x)\,x - \alpha_t(y)\,y \right|
\end{equation}
\begin{equation}
    \textbf{G}_{\beta, t} \coloneqq \frac{1}{2\, \textbf{E}_{p_{t}}\left[ \beta_t(x) \right]   } \int dx dy\, p_{t}(x) p_{t}(y) \, |\beta_t(x) - \beta_t(y)|
\end{equation}
With this decomposition, we can write the change in the Gini coefficient over time as:
\begin{equation}
\label{gini_evolution}
    \textbf{G}_{x, t+1}  =  \textbf{G}_{\alpha x, t} + \frac{\varrho_t}{1+\varrho_t}  \left( \textbf{G}_{\beta, t} - \textbf{G}_{\alpha x, t} \right)+ \textbf{G}_{\Gamma, t}
\end{equation}
Where we introduced the ratio:
\begin{equation}
\varrho_t \coloneqq \frac{\textbf{E}_{p_t}\left[  \beta_t(x) \right]}{\textbf{E}_{p_t}\left[  \alpha_t(x)\, x \right]}
\end{equation}
From equation \eqref{gini_evolution} we see that the condition for $\textbf{G}_{x, t}$ to stop increasing is given by:
\begin{equation}
\label{eq:convergence_condition}
    \frac{\varrho_t}{1 + \varrho_t}  \geq  \frac{  \left( \textbf{G}_{\alpha x, t}  - \textbf{G}_{x, t}  \right)+  \textbf{G}_{\Gamma, t}  }{\left(  \textbf{G}_{\alpha x, t} - \textbf{G}_{\beta, t} \right)}
\end{equation}
The last inequality implies that, at any level of concentration as measured by the Gini coefficient, it is possible to stop its growth by sufficiently increasing the ratio between the average of the additive component of the growth process and that of the multiplicative one.  
In order for the Gini index not to saturate, the coefficient $\varrho_t$ cannot become arbitrarily small. 
In addition, as long as $\textbf{G}_{\beta, t}+\textbf{G}_{\Gamma, t}<1$, this is also a sufficient condition to ensure a bounded wealth inequality, as we show in the following.
In appendix \ref{app:alpha_growth} we prove the inequality:
\begin{equation}
\label{eq:financial_gini_bound}
    \textbf{G}_{\alpha x, t} \geq \textbf{G}_{x,t} 
\end{equation}
for any non-decreasing function $\alpha_t(x)$.
As a consequence, the only regulating term in equation \eqref{gini_evolution} is $\left(\textbf{G}_{\beta, t} - \textbf{G}_{\alpha x, t}\right)$. If it is not negative, concentration is forced to increase in time, and the Gini index saturates. For this reason we limit our analysis to the case $\textbf{G}_{\beta, t} < \textbf{G}_{\alpha x, t}$. 
This implies that for any value of $\varrho_t$ there is a lower bound for the evolution of $\textbf{G}_{x, t}$, given by:
\begin{equation}
    \textbf{G}_{x, t+1}  \geq \textbf{G}_{\beta, t} + \textbf{G}_{\Gamma, t},
\end{equation}
hence we will assume $\textbf{G}_{\beta, t}+\textbf{G}_{\Gamma, t} \leq \textbf{G}_{x, t}$ (otherwise we would a trivial growth of the Gini coefficient).
In the limit $\varrho_t \to 0$, i.e. when the ratio between the additive term $\textbf{E}_{p_t}\left[  \beta_t(x) \right]$ and the multiplicative one $\textbf{E}_{p_t}\left[  \alpha_t(x)\, x \right]$ goes to zero, equation \eqref{gini_evolution} becomes:
\begin{equation}
    \textbf{G}_{x, t+1} = \textbf{G}_{\alpha x, t} + \textbf{G}_{\Gamma, t} 
\end{equation}
Assuming the validity of hypothesis \ref{hyp:positive_dispersion}, this can be further reduced to:
\begin{equation}
\label{growth_case_1}
    \textbf{G}_{x, t+1} \geq \textbf{G}_{x, t} + \frac{\Omega\left( \Gamma_t\right)}{g_t}  \kappa\,  \textbf{P}_{t}\left( x \geq \kappa\, \textbf{E}_{p_{t}}\left[ x \right] \right)^2
\end{equation}
where we introduced the lower bound for $\mathbf{E}_{p_t}\left[ F_t(x,y) \right]$ and the average growth factor:
\begin{equation}
g_t = \frac{\textbf{E}_{p_t}\left[   \gamma_t(x) \right]}{\textbf{E}_{p_t}\left[  x \right]}.
\end{equation}
The last term on the right hand side of equation \eqref{growth_case_1} is positive unless the Gini index is already at its maximum. For this reason, in this case the inequality is bound to keep increasing.

On the other hand, the term $\left( \textbf{G}_{\alpha x, t}  - \textbf{G}_{x, t}  \right)$ on the right hand side of inequality \eqref{eq:convergence_condition} becomes null when the Gini index tends to one (we have $\textbf{G}_{x, t} < \textbf{G}_{\alpha x, t} < 1$); 
in addition the upper bound \eqref{eq:f_upper_bound} implies:\footnote{A saturating Gini index implies a null limit for $P_t\left( x\geq y\right)$, for any $y$. Hence the integral defining $\textbf{G}_{\Gamma, t}$ becomes arbitrarily small for $\textbf{G}_{x, t} \to 1$.}
\begin{equation}
    \lim_{\textbf{G}_{x, t} \to 1} \textbf{G}_{\Gamma, t} = 0
\end{equation}
These two facts imply that the right hand side of condition \eqref{eq:convergence_condition} becomes arbitrarily small when the Gini coefficient grows close to one. As a consequence, any positive value of $\varrho_t$ will stop the growth of inequality at a finite level. 
Hence a positive value of $\varrho_t$ is both a necessary and sufficient condition to have a bounded concentration of wealth.
In other words, in order to have a bounded wealth inequality, the average of the additive term in wealth growth $\textbf{E}_{p_t}\left[  \beta_t(x) \right]$ must grow at least proportionally to the multiplicative term $\textbf{E}_{p_t}\left[  \alpha_t(x)\, x \right]$.
Finally when $\varrho_t \to \infty$ the evolution of the Gini index reduces to:
\begin{equation}
\label{growth_case_2}
    \textbf{G}_{x, t}  = \textbf{G}_{\beta, t} + \textbf{G}_{\Gamma, t} 
\end{equation}
and in this case the dynamics of the Gini index is driven by the additive term $\beta_t(x)$ plus a small contribution from the dispersion of the wealth growth evolution.

We obtained a necessary and sufficient condition on the average of the non-multiplicative part of the wealth evolution for the inequality not to increase. 
This is the main result of the paper, and in the next section we will discuss the interpretation in terms of more easily identifiable economic quantities. 



\section{Interpretation in terms of specific models}
\label{sec:applications}
In this section we want to show how the general derivation of the previous section translates in term of simple examples of wealth growth models. 
The previous treatment is much more general than any particular model, hence obviously the conclusions we can derive from it are only a portion of the ones obtained by solving the model. 
Nonetheless, it can be useful to see how the elements of our derivation are mapped onto terms with clear economical interpretations. 
All the models we analyze here are of the form:
\begin{equation}
\label{eq:example_model}
    \hat{x}_{t+1}^i = e^{\hat{r}_t^i}\, \left( x_t^i - c_t^i \right) + \hat{y}_t^i
\end{equation}
where $i$ is the index identifying each agent, $x_t$ it the wealth value at time $t$, $\hat{r}_t$ the stochastic returns on wealth, $\hat{y}_t$ the stochastic labor income term, $c_t$ the amount of wealth consumed in the period $t$ by each agent.
These models are among the building blocks of most of the literature on wealth inequality (\cite{gabaix2016dynamics, cagetti2008wealth, benhabib2015wealth, benhabib2018skewed}), and various assumptions about the processes $\hat{r}_t$, $\hat{y}_t$ and about the consumption choices $c_t$ were shown to imply interestingly different types of dynamics. 
Equation \eqref{eq:example_model} can be mapped to equation \eqref{general_stochastic_growth} with the identifications:
\begin{equation}
    \hat{A}_t^i(x_t, \xi_t) = e^{\hat{r}_t^i}\, \left( 1 - \frac{c_t^i}{x_t^i} \right) \quad , \quad \hat{B}_t^i(x_t, \xi_t) = \hat{y}_t^i
\end{equation}
where the meaning of the parameters $\xi_t^i$ depends on the definitions of $c_t^i$ and $\hat{y}_t^i$.
These identifications allow us to interpret the multiplicative term $\hat{A}_t^i(x_t, \xi_t)$ as the invested wealth, and the additive term $\hat{B}_t^i(x_t, \xi_t)$ as labor income. 
We consider for the stochastic returns $\hat{r}_t$ a memory-less Gaussian variable, with mean $\mu_r$ and standard deviation $\sigma_r$:
\begin{equation}
    \hat{r}_t \sim N\left(\mu_r , \sigma_r\right)
\end{equation}
Regarding the labor income terms, we assume for them to growth at a rate $\dot{\mu}_y$:
\begin{equation}
    y_t^i = e^{\dot{\mu}_y}\,y_{t-1}^i 
\end{equation}
and we will see that the value of this growth rate is crucial for the existence of an equilibrium in wealth inequality. 
The stochastic nature of $\hat{y}_{t}$ is assumed to derive only from a non-trivial distribution in the population of agents, and not from uncertainty in the time evolution. 

The last component to be modeled are consumption choices $c_t$. They are usually described as a maximization process, in which each agent looks to maximize a utility function, given its wealth and expectations about the future. 
We will analyze the behaviour of three different models in terms of these consumption choices; in the first one (which we will denote as \textbf{M1}) we will assume negligible consumption: $\frac{c_t}{x_t} \ll 1$. While this assumption is not realistic in most situation, it can be useful to explore the dynamics of inequality in a regime of unchecked wealth growth.
The evolution of wealth reduces in this case to:
 \begin{equation}
     \hat{x}_{t+1}^i = e^{\hat{r}_t^i}\, x_t^i + \hat{y}_t^i
 \end{equation}
 and we will see the crucial role of the labor income term $\hat{y}_t$ in regulating the growth of inequality. 
 For the second model (which we will denote as \textbf{M2}) we will assume consumption choices generated by the maximization of a CRRA utility function, of the form:
\begin{equation}
    u(c_t) = \frac{c_t^{1-\gamma}}{1-\gamma} \quad , \quad \gamma > 0. 
\end{equation}
We will see that in this case the assumption of the model are such that it always reaches equilibrium, both in terms of an asymptotic distribution of wealth and in terms of inequality. 
 Finally we will exploit the fact, derived in section \ref{sec:gini_evolution}, that the only relevant way in which consumption choices influence the evolution of inequality is through the average $\alpha_t(x)$ (and, less critically, through their dispersion in $\textbf{G}_{\Gamma, t}$). For this reason, the third model we will study (denoted as \textbf{M3}) is one in which consumption choices are not derived from a utility maximization, but described at the level of average consumption over all the agents with a given level of wealth $x_t = x$. We denote this (deterministic) aggregate function as $c_t(x)$, and assume its form to be:
 \begin{equation}
     c_t(x) = \frac{x}{1 + \left(\frac{x}{\mu_c}\right)^{\sigma_c}}
 \end{equation}
for some coefficients $\mu_c > 0$, $\sigma_c \in \left(0, 1\right)$. Assuming the consumption of each agent to be described by this average function, the only error we are committing is an underestimation of the dispersion term $\textbf{G}_{\Gamma, t}$.

For all the three models presented it is trivial to verify the validity of hypotheses \ref{hyp:positivity_coefficients}, \ref{hyp:positive_dispersion}, and in all the models the inequality reaches an equilibrium only when the average labor income remains proportional to the average invested wealth. 

\subsection{Model \textbf{M1}}
With trivial consumption choices, this model reduces to a combination of a multiplicative evolution given by financial returns, and an additive term given by labor incomes. 
As anticipated, this model is not expected to be a realistic description of the dynamics of the bulk distribution of wealth. 
For agents with very high wealth, however, we can expect the consumption-to-wealth ratio to be small, such that in equation \eqref{eq:example_model} the consumption term can be neglected.\footnote{This is not true in all models: in model \textbf{M2}, for example, we will see that the consumption term is critical for the existence of an equilibrium distribution of wealth.} 
With $c_t = 0$ the wealth evolution reduces to:
\begin{equation}
\label{eq:m1_evo}
    \hat{x}_{t+1} = e^{\hat{r}_t}\, x_t + \hat{y}_t \quad , \quad \hat{r}_t \sim N\left(\mu_r , \sigma_r\right)
\end{equation}
and we assume $\sigma_r > 0$. 
It is easy to derive the identities:
\begin{equation}
    \textbf{E}_{p_t}\left[ \alpha_t(x)\,x \right] =  \textbf{E}\left[ e^{\hat{r}_t} \right] \textbf{E}_{p_t}\left[ x \right] \quad , \quad \textbf{G}_{\alpha x, t} = \textbf{G}_{x, t}.
\end{equation}
In particular we have:
\begin{equation}
    \alpha_t(x) =\textbf{E}\left[ e^{\hat{r}_t} \right].
\end{equation}
The labor income is assumed to follow the dynamics:
\begin{equation}
    y_t^i = e^{\dot{\mu}_y}\, y_{t-1}^i
\end{equation}
The process $\hat{y}_t$, though independent of $\hat{x}_t$ in its updates, has a memory, and its realizations influence future values of $\hat{x}_t$. This implies\footnote{See appendix \ref{app:positive_beta} for a detailed discussion.} that higher values of $\hat{y}_t$ are correlated with higher values of $\hat{x}_t$, and the validity of hypothesis \ref{hyp:positivity_coefficients}. Hypothesis \ref{hyp:positive_dispersion} is trivially satisfied as long as $\sigma_r > 0$.

It is easy to see that the long term evolution of wealth inequality in this model depends only on the relative magnitude of the growth rates of the financial wealth and labor incomes. 
With the description of the labor income terms $\hat{y}_t$ as positive \emph{i.i.d.} random variables (hence $\dot{\mu}_y = 0$), equation \eqref{eq:m1_evo} describes a Kesten process (\cite{champernowne1953model, kesten1973random, benhabib2015wealth}), and the dynamics of the model can be characterized in a rigorous manner. 
In particular, it has been proven in \cite{kesten1973random} that a necessary condition for the existence of a limiting equilibrium distribution for the process \eqref{eq:m1_evo} is $\textbf{E}\left[ \hat{r}_t \right] < 0$. 
When this is the case, the boundedness of the Gini coefficient is linked to the existence of the mean of the distribution. 
With the following notation:
\begin{equation}
    e^{\overline{r}} \coloneqq \textbf{E}\left[ e^{\hat{r}_t} \right] 
\end{equation}
we can describe the average evolution of wealth in this model as:
\begin{equation}
\label{eq:exponential_growth}
    \textbf{E}\left[ \left. \hat{x}_{t+n} \right| x_t = x, \, y_t = y \right] = e^{n\,\overline{r}} \, x  + y\, \frac{1 - e^{ n\, \overline{r}} }{1  - e^{\overline{r}} } 
\end{equation}
From the last expression it is easy to see that the condition for the existence of an average coincides with the one of a finite limit for the ratio $\varrho_t = \frac{\textbf{E}_{p_t}\left[  \beta_t(x) \right]}{\textbf{E}_{p_t}\left[  \alpha_t(x)\, x \right]}$: 
for $\overline{r} < 0$ expression \eqref{eq:exponential_growth} has the finite limit:
\begin{equation}
    \lim_{n\to \infty} \textbf{E}\left[ \left. \hat{x}_{t+n} \right| x_t = x, \, y_t = y  \right] =  \frac{ y }{1 - e^{\overline{r}}} 
\end{equation}
Averaging over the values of $x$ and $y$ we find the (finite) average value of the long-term distribution of wealth:
\begin{equation}
\label{eq:m1_limit}
    \textbf{E}_{p_t}\left[ x \right] = \frac{ \overline{y} }{1 - e^{\overline{r}}} 
\end{equation}
where we defined the average labor income:
\begin{equation}
    \overline{y}\coloneqq \textbf{E}_{p_t}\left[ \beta_t(x) \right].
\end{equation}
Equation \eqref{eq:m1_limit} implies in addition: 
\begin{equation}
\lim_{t \to \infty} \varrho_t = e^{|\overline{r}|} - 1.
\end{equation}
Hence we find that the existence of an average for the long-time wealth distribution coincides with a finite upper bound in the concentration of wealth. 
If on the other hand $\overline{r} \geq 0$ (while $\textbf{E}\left[ \hat{r}_t \right] < 0$), the conditional average wealth \eqref{eq:exponential_growth} diverges for every value of the starting wealth $x$, we do not have a definite average for the long-time wealth distribution, and the ratio:
\begin{equation}
    \varrho_t = \frac{\textbf{E}_{p_t}\left[  \beta_t(x) \right]}{\textbf{E}_{p_t}\left[  \alpha_t(x)\, x \right]} 
\end{equation}
tends to zero, implying an unbounded growth of wealth inequality.

Our analysis, however, is more general than the available results for Kesten processes: the condition $\textbf{E}\left[ \hat{r}_t \right] < 0$ (and the existence of an asymptotic equilibrium distribution) is not necessary to describe the evolution of inequality in model \eqref{eq:m1_evo}. 
For $\textbf{E}\left[ \hat{r}_t \right] \geq 0$ we don't have any equilibrium distribution, but we can nevertheless apply the results of section \ref{sec:gini_evolution}. 
When the labor income growth rate $\dot{\mu}_y$ is greater than zero, the evolution of expected wealth becomes:\footnote{We start the evolution at $t=0$ because for $\textbf{E}\left[ \hat{r}_t \right] \geq 0$ the system can have infinite memory, so that all its history becomes relevant.}
\begin{equation}
\label{eq:general_conditional_evolution_m1}
    \textbf{E}\left[ \left. \hat{x}_{t} \right| x_0 = x ,\, y_{0}=y \right] = e^{\overline{r}\,t} \, x + e^{\dot{\mu}_y\, ( t -1 )} \, y \, \frac{1 - e^{ \left( \overline{r} - \dot{\mu}_y \right)\, t } }{  1  - e^{ \left( \overline{r} - \dot{\mu}_y \right)  } } 
\end{equation}
and the average labor income becomes:
\begin{equation}
    \textbf{E}_{p_t}\left[ \beta_t(x) \right] = e^{\dot{\mu}_y\,(t + 1)}\, \overline{y}
\end{equation}
where we used $\overline{y}$ for the average labor income at time $t=0$.
Taking the ratio of the average of the two quantities over the whole population, we find:
\begin{equation}
    \varrho_t = \overline{y}  \,  e^{\left( \dot{\mu}_y - \overline{r} \right) \,( t + 1)}\, \left[  \overline{x} +  \, \overline{y} \, \frac{e^{ \left( \dot{\mu}_y - \overline{r}\right)\, t} - 1}{  1  - e^{ \left( \overline{r} - \dot{\mu}_y \right)  } } \right]^{-1}  
\end{equation}
where $\overline{x}$ is the initial average wealth. 
We can see that for $\overline{r} \geq \dot{\mu}_y$ we obtain a null limit for $\varrho_t$, which implies a saturation of the Gini index. 
For $\overline{r} < \dot{\mu}_y$, on the other hand, we obtain the finite limit:
\begin{equation}
    \lim_{t \to \infty} \varrho_t = e^{\left( \dot{\mu}_y - \overline{r} \right) } - 1
\end{equation}
This implies a Gini coefficient less than one, and a stable concentration of wealth. Note that in this case for either $\dot{\mu}_y > 0$ or $\overline{r} \geq 0$ we don't have a definite long-term limit for the average wealth.

\subsection{Model \textbf{M2}}
\label{sec:m2}
In this section we want to describe a model which takes into account consumption choices of the agents. In particular, we describe consumption as derived from the maximization of some utility function for each agent. Here we choose a CRRA utility function, which is among the most studied in the economics literature (\cite{huggett1993risk, aiyagari1994uninsured, gabaix2016dynamics, xavier2021wealth}). The underlying hypothesis of this kind of models is that each agent chooses the amount of wealth to consume in order to maximize the function:
\begin{equation}
\label{eq:utility}
    u(c_t) = \frac{c_t^{1-\gamma}}{1-\gamma} \quad , \quad \gamma >0 
\end{equation}
for every period $t$, subject to the constraint $c_t \leq x_t$. Here $c_t$ is the amount of wealth consumed by a given agent at time $t$. 
The optimal consumption choices are obtained solving the multi-period optimization problem described by the Bellman equation (\cite{bellman1952theory, bellman1966dynamic}):
\begin{equation}
\label{eq:bellman_equation}
    V(x_t, \xi_t) = \max_{c_t \leq x_t} \left\{ u(c_t) + \beta\,\textbf{E}\left[ V\left( \hat{x}_{t+1}(x_t, c_t, \xi_t),\, \hat{\xi}_{t+1} \right) \right] \right\}
\end{equation}
The function $V(x_t, \xi_t)$ is the so-called value function of the problem, representing the expectation value of all the future consumption utilities of the optimizing agent.
The parameters $\xi_t$ represent all the external factors influencing the expectation value of $\hat{x}_{t+1}$ (in our case they represent primarily past values of the labor income), and the discount factor $\beta \in (0,1)$ takes into account the importance of future consumption relative to the present.

We assume again for the process $\hat{r}_t$ to be independent of all other variables in the model, and for the labor income the evolution:
\begin{equation}
    y_t^i = y_{t-1}^i.
\end{equation}
More generally, whenever $\hat{y}_t^i = \hat{z}_t^i\,y_{t-1}^i$ for some positive random variable $\hat{z}_t^i$, we can write the Bellman equation as:
\begin{equation}
    V(x_t, y_{t-1}) = \max_{c_t \leq x_t} \left\{ u(c_t) + \beta\,\textbf{E}\left[ V\left( e^{\hat{r}_t}\,\left( x_t - c_t \right) + \hat{z}_t\, y_{t-1} , \,\hat{z}_t \, y_{t-1} \right) \right] \right\}
\end{equation}
and we can see that the value function depends on the past realization of labor income together with the value of wealth (and the parameters of the distributions of both quantities). 
In appendix \ref{app:consumption} we prove two useful properties of the solution of the dynamic optimization problem stated above, derived from the CRRA form of the utility function \eqref{eq:utility}.
First, the value function $V(x_t, y_{t-1})$ can be written in the form:\footnote{This is a generalization of the homogeneity property shown for example in \cite{toda2017discrete}. See also \cite{carroll2004theoretical} for a similar derivation.}
\begin{equation}
    V(x_t, y_{t-1}) = \tilde{V}\left( \frac{x_t}{y_{t-1}}\right) \, x_t^{1-\gamma}
\end{equation}
The Bellman equation is thus equivalent to:
\begin{equation}
    \tilde{V}\left(\frac{x_t}{y_{t-1}} \right) = \max_{\tilde{c}_t \leq 1} \left\{ u(\tilde{c}_t) + \beta\,\textbf{E}\left[ \left( e^{\hat{r}_t}\,\left( 1 - \tilde{c}_t \right) + \hat{z}_t\,\frac{y_{t-1}}{x_t} \right)^{1-\gamma} \tilde{V}\left( \frac{\hat{x}_{t+1}}{\hat{y}_t}\left( \tilde{c}_t  \right)\right) \right] \right\}
\end{equation}
with:
\begin{equation}
     \frac{\hat{x}_{t+1}}{\hat{y}_t}\left( \tilde{c}_t  \right) = \frac{e^{\hat{r}_t}}{\hat{z}_t}\,\left( 1 - \tilde{c}_t \right)\, \left(\frac{x_t}{y_{t-1}} \right) + 1.
\end{equation}
Expressed in this form, the Bellman equation implies:
\begin{equation}
    \tilde{c}_t = \tilde{c}_t\left( \frac{x_t}{y_{t-1}} \right),
\end{equation}
and the same property holds for the fraction of saved wealth $s_t$:
\begin{equation}
    s_t\left( \frac{x_t}{y_{t-1}} \right) \coloneqq 1 - \tilde{c}_t\left( \frac{x_t}{y_{t-1}} \right)
\end{equation}
From these facts we derive that for each agent the conditional evolution of wealth can be written as:
\begin{equation}
\label{eq:independent_evolution}
     \hat{x}_{t+1} = \hat{g}_t\left( \frac{x_t}{y_{t-1}} \right) x_t  
\end{equation}
with:
\begin{equation}
    \hat{g}_t\left( \frac{x}{y} \right) \coloneqq s_t\left( \frac{x}{y} \right)  e^{\hat{r}_t}  + \frac{y}{x} 
\end{equation}
This implies that the evolution of the ratio $\frac{x}{y}$ depends only on the ratio itself, and not on the separate values of $x$ and $y$.
A second useful fact shown in appendix \ref{app:consumption} is the asymptotic property of the saving function:
\begin{equation}
\label{eq:asymptotic_saving}
    \lim_{x_t \to \infty} s_t\left( \frac{x_t}{y_{t-1}} \right) = \eta < 1.
\end{equation}
In order for the dynamics of the model to reach an asymptotic equilibrium distribution, it is necessary to have:\footnote{Given a generic evolution map $\psi_{\theta}(x)$, a condition for the existence of an equilibrium distribution is given by (see \cite{diaconis1999iterated}):
\begin{equation}
    \textbf{E}\left[ \log L_{\theta}\right] <0,
\end{equation}
where $L_{\theta}$ is the Lipschitz constant of the map $\psi_{\theta}$:
\begin{equation}
    L_{\theta} \coloneqq \sup_{x\neq y} \frac{\left| \psi_{\theta}(x)  - \psi_{\theta}(y) \right|}{\left| x-y \right|}.
\end{equation}
In our case, using the convexity of the saved wealth function $(x_t - c_t)$ (which in turn is derived from the concavity of the consumption function), this condition reduces to equation \eqref{eq:log_contraction}.
}
\begin{equation}
    \label{eq:log_contraction}
    \log \eta + \textbf{E}\left[ \hat{r}_t \right] < 0. 
\end{equation}
With this hypothesis the dynamics is on average contracting, and the equilibrium distribution generated by the process \eqref{eq:independent_evolution} does not depend on its initial value of the ratio $\frac{x}{y}$ (see \cite{diaconis1999iterated, mirek2011heavy} for details). In other words, after a sufficient time we obtain for the ratio $\frac{x}{y}$ an equilibrium distribution independent on the separate values of $x$ and $y$. 
Leveraging these facts, in appendix \ref{app:positive_beta} we show that after a sufficient time the dynamics just described implies:
\begin{equation}
    \frac{d}{d x} \alpha_t(x) = \textbf{E}\left[ e^{\hat{r}_t} \right] \frac{d}{d x} \textbf{E}\left[ \left. s_t\left( \frac{x_t}{y_{t-1}}\right)  \right| x_t = x  \right] \geq 0
\end{equation}
\begin{equation}
    \frac{d}{d x} \beta_t(x) \geq 0
\end{equation}
\begin{equation}
    \textbf{E}\left[ \left. s_t\left( \frac{x_t}{y_{t-1}}\right)  \right| x_t = x  \right]  > 0 
\end{equation}
such that all the hypotheses of our derivation are valid.
We found that any equilibrium quantile of wealth is a fixed multiple of the labor income for each agent, and any positive shock in wealth not related to labor income (e.g. a large positive fluctuation in returns $\hat{r}_t$) is consumed in a finite time. This leaves labor income as the main driver in the evolution of inequality. 
In particular, equation \eqref{eq:independent_evolution} implies:
\begin{equation}
    \lim_{t \to \infty}\varrho_t = \left( \textbf{E}\left[ e^{\hat{r}_t} \right] \textbf{E}\left[  s_t\left( \frac{x_t}{y_{t-1}}\right) \frac{x_t}{y_{t-1}} \right] \right)^{-1}.
\end{equation} 
Given the finite limit \eqref{eq:asymptotic_saving}, $\varrho_t$ is different from zero if and only if we have a finite average wealth, similarly to what we found in model \textbf{M1}. Hence a finite average wealth implies a finite limit for wealth concentration. 

In order to ensure the existence of a solution for the Bellman equation \eqref{eq:bellman_equation}, often the stronger assumption is made (see \cite{benhabib2015wealth} for a full analytical description of the model): 
\begin{equation}
\label{eq:contraction}
    \eta <   \textbf{E}\left[ e^{\hat{r}_t} \right]^{-1}.
\end{equation} 
As shown in appendix \ref{app:positive_beta}, the last condition implies a finite average wealth, hence in this case a bounded concentration of wealth is effectively assumed in the dynamics of the model.

\subsection{Model \textbf{M3}}
One of the defining features of model \textbf{M2} is the finite memory, which conflicts with the observed long persistence of wealth (see for example \cite{barone2016intergenerational}). In this section we exploit the possibility of describing wealth dynamics at the aggregate level, and introduce an evolution process defined by an average consumption function. 
This model can be seen as a combination of \textbf{M1} and \textbf{M2}: for low values of wealth it is contracting (meaning that any fluctuation above the equilibrium wealth is consumed in a finite time), and its dynamics resembles the one generated by model \textbf{M2}. Above a certain wealth threshold, however, wealth tends to grow indefinitely, and wealth inequality can escape equilibrium, similarly to what we saw in model \textbf{M1}. The equilibrium of this system is \emph{metastable}. 
The dynamics will oscillate around its equilibrium point for a long time, but eventually in any finite population some agent will start accumulating wealth in an unbounded way, driving wealth inequality to a divergence.  
The main difference between models \textbf{M3} and \textbf{M2} is the large-wealth behaviour of the saving function $s_t$. 
In the most common specification of model \textbf{M2} it is assumed that $\alpha_t(x) =  \textbf{E}\left[ e^{\hat{r}_t} \right] \textbf{E}\left[ \left. s_t\left( \frac{x_t}{y_{t-1}}\right)  \right| x_t = x  \right] <1$ for any $x$. 
The more general consumption function assumed in model \textbf{M3} does not imply $\alpha_t(x)<1$, and its dynamics can encompass both the contracting and expansive regimes. In particular, we have a critical value of wealth $x_{**}$ such that above it the dynamic will switch from a contracting trend similar to the one found in model \textbf{M2} to an expansive trend similar to the one of model \textbf{M1}.  
The dynamics of this model, given wealth $x_t$, is described by the evolution law:
\begin{equation}
    \hat{x}_{t+1} = s_t\left( x_{t} \right)  e^{\hat{r}_t} \, x_t + \hat{y}_t
\end{equation}
with:
\begin{equation}
\label{eq:savings}
    s_t(x) \coloneqq \left( 1 - \frac{c_t(x)}{x}\right) \quad , \quad c_t(x) \coloneqq \frac{x}{1 + \left(\frac{x}{\mu_c}\right)^{\sigma_c}}
\end{equation}
for some coefficients $\mu_c > 0$, $\sigma_c \in \left(0, 1\right)$.
In addition, for the returns on wealth we assume as in the previous sections a normal distribution:
\begin{equation}
    \hat{r}_t \sim N\left(\mu_r , \sigma_r\right)
\end{equation}
and the labor income terms are assumed to be fixed for each agent: $y^i_{t} = y^i_{t-1}$, with log-normal distribution over the population:
\begin{equation}
    \hat{y}_t \sim LN\left(\mu_{y}, \sigma_y \right).   
\end{equation}
The function $\alpha_t(x) = \textbf{E}\left[ e^{\hat{r}_t} \right] s_t(x)$ is positive and increasing, and it can easily be checked that also $\beta_t(x)$ is non-decreasing. Hypothesis \ref{hyp:positivity_coefficients} is thus satisfied. The saving function tends to one for large values of wealth, which implies:
\begin{equation}
    \lim_{x\to \infty}  \textbf{E}\left[ e^{\hat{r}_t} \right] s_t(x) = \textbf{E}\left[ e^{\hat{r}_t} \right] > 0
\end{equation}
and the validity of hypothesis \ref{hyp:positive_dispersion}. It is worth noting that if we assume the function $c_t(x)$ defined in \eqref{eq:savings} to correctly represent the average consumption for agents with wealth $x$, we can describe each agent as following this rule without introducing any bias in the aggregate behaviour of wealth. 
The only error we are introducing is an under-estimation of the dispersion of the wealth growth process. 

While for model \textbf{M1} returns on wealth growing faster than labor incomes imply an unbounded concentration of wealth, here we can find at least a section of the population in a non-trivial equilibrium. 
We can write the average wealth growth equation as:
\begin{equation}
     \textbf{E}\left[ \left. \hat{x}_{t+1} \right| x_t=x , \,y_{t-1} = y  \right] = h\left( x,y \right) x 
\end{equation}
with
\begin{equation}
\label{eq:general_stability}
    h\left( x, \, y \right) = e^{\overline{r}}\frac{x^{\sigma_c}}{\mu_c^{\sigma_c} + x^{\sigma_c}} + \frac{y}{x}
\end{equation}
where as in the last sections we used the notation $e^{\overline{r}} \coloneqq \textbf{E}\left[ e^{\hat{r}_t} \right]$.
It is easy to show\footnote{Expressing the wealth $x$ in units of $\mu_c$, equation \eqref{eq:general_stability} becomes:
\[
    h\left( \zeta \mu_c, \, y \right) = \frac{ e^{\overline{r}} }{1 + \zeta^{-\sigma_c}} + \frac{y}{\mu_c}\frac{1}{\zeta} 
\]
For a small enough ratio $y/\mu_c$ the minimum of this expression becomes arbitrarily small, and given the limits of the expression for $\zeta \to 0$ and $\zeta \to \infty$, in at least one of the solutions of the equation $h\left( \zeta \mu_c, \,y \right) = 1$ the function is decreasing, which implies the stability of equilibrium.} that for a small enough ratio $\frac{y}{\mu_c}$ this equation implies a stable equilibrium point $x_*(y)$, such that for small values of $\epsilon$:
\begin{equation}
     h\left( x_*(y), \, y  \right) = 1 
\end{equation}
\begin{equation}
    h\left( x_*(y) + \epsilon , \, y  \right) < 1 \quad , \quad h\left( x_*(y) - \epsilon , \, y \right) > 1
\end{equation}
Denoting as $x_{**}(y)$ the upper bound of the attraction basin of this equilibrium point, this implies that all agents with labor income $y$ and wealth $x<x_{**}(y)$ are on average driven by the dynamics to the stable equilibrium $x_*(y)$. If the initial conditions of the model are such that all agents are in this attraction basin, the (average) dynamics is similar to the one of model \textbf{M2}, and wealth concentration remains bounded for a long time. Whenever some agents' wealth fluctuates to a value higher than $x_{**}(y)$, however, on average it will keep increasing, and with it the averages of $x$ and $\alpha_t(x)\,x$. 
This implies the limit:
\begin{equation}
    \lim_{t \to \infty} \varrho_t = 0, 
\end{equation}
and an unbounded increase in wealth concentration.
This provides an example of model which can appear in equilibrium for a very long time, but inevitably transitions to a state of ever-increasing inequality. In addition, it shows the advantage of using methods applicable both with and without an equilibrium asymptotic distribution.

\section{Conclusions}\label{sec:concl}
In this paper we proved a short series of results regarding the evolution of wealth inequality in models with multiplicative wealth growth and additive labor income terms.
In particular we prove, in a general class of models, that a necessary and sufficient condition to avoid an unbounded growth of wealth concentration is the existence of a lower bound on the ratio between the additive (labor income) term and the multiplicative (invested wealth) term. We find that a decrease in this ratio corresponds to an increase in wealth concentration, and in particular that whenever this ratio tends to zero wealth inequality is bound to diverge. 
In this form, our result is similar to the one described in \cite{piketty2014inequality} analyzing historical data, and found again in \cite{piketty2015wealth} in the context of a specific model.   
An advantage of the derivation presented here is the independence from the concept of an equilibrium wealth distribution, which is a necessary ingredient in the study of inequality in most of the economic literature. 
While the importance of labor income terms to guarantee the existence of an equilibrium distribution is pointed out in many models (see \cite{champernowne1953model, kesten1973random, benhabib2015wealth, gabaix2016dynamics}), to our knowledge a general description is not available in the literature. In addition, we show explicitly the general independence of the existence of an equilibrium asymptotic wealth distribution and the existence of an equilibrium in the level of inequality.


\printbibliography 
 
\clearpage
 
\appendix

\section{Gini index decomposition}
\label{app:gini_decomposition}
We want to prove the general decomposition of the Gini index:
\begin{equation}
    \textbf{G}_{x, t+1} = \textbf{G}_{\gamma, t} + \textbf{G}_{\Gamma, t} 
\end{equation}
with:
\begin{equation}
    \textbf{G}_{x,t} \coloneqq \frac{1}{2\,\textbf{E}_{p_t}\left[ x \right]} \int dx dy\, p_t(x) p_t(y)\, |x-y|
\end{equation}
\begin{equation}
    \textbf{G}_{\gamma, t} \coloneqq \frac{1}{2\, \textbf{E}_{p_{t}}\left[ \gamma_t(x) \right]   } \int dx dy\, p_{t}(x) p_{t}(y) \, \left|\gamma_t(x) - \gamma_t(y) \right|
\end{equation}
\begin{equation}
    \textbf{G}_{\Gamma, t}  \coloneqq \frac{\mathbf{E}_{p_t}\left[F_t\left( x, y\right) \right]}{\textbf{E}_{p_{t}}\left[ \gamma_t(x) \right]  }
\end{equation}
\begin{align}
    F_t(x,y) \coloneqq  \, \mathbb{1}\left( x \geq y \right)  & \,\int_{y'>x'} dx'dy'\,w_t(x'|x) w_t(y'|y) \,|x' - y'| \, + \\
    & + \mathbb{1}\left( y > x \right)\, \int_{x'>y'}  dx'dy'\,w_t(x'|x) w_t(y'|y) \,|x' - y'|
\end{align}
To perform this task, we exploit the trivial identity regarding the absolute value of a number:
\begin{equation}
    |x-y| = (x-y) + 2\, (y-x)\mathbb{1}\left( y \geq x \right)
\end{equation}
to write:
\begin{align}
    \textbf{G}_{x, t+1} = \frac{1}{2\,\textbf{E}_{p_t}\left[ \gamma_t(x) \right]} & \int dx dy\, p_t(x) p_t(y)\, \int dx' dy'\, w_t(x'|x) w_t(y'|y) \,|x' - y'| = \\
    = \frac{1}{\textbf{E}_{p_t} \left[ \gamma_t(x) \right]}  & \int_{x>y} dx dy\, p_t(x) p_t(y)\, \int dx' dy'\, w_t(x'|x) w_t(y'|y) \,|x' - y'| = \\
    =  \frac{1}{\textbf{E}_{p_t}\left[ \gamma_t(x) \right]} & \int_{x>y} dx dy\, p_t(x) p_t(y)\, \int dx' dy'\, w_t(x'|x) w_t(y'|y) \,\left( x' - y' \right) + \\
    + & \frac{2}{\textbf{E}_{p_t}\left[ \gamma_t(x) \right]} \int_{x>y} dx dy\, p_t(x) p_t(y)\, \int_{y'>x'} dx' dy'\, w_t(x'|x) w_t(y'|y) \,\left(y'-x' \right) = \\
    = \frac{1}{\textbf{E}_{p_t}\left[ \gamma_t(x) \right]} & \int_{x>y} dx dy\, p_t(x) p_t(y)\, \left( \gamma_t(x) - \gamma_t(y) \right) + \\
     + & \frac{2}{\textbf{E}_{p_t}\left[ \gamma_t(x) \right]} \int_{x>y} dx dy\, p_t(x) p_t(y)\, \int_{y'>x'} dx' dy'\, w_t(x'|x) w_t(y'|y) \,\left(y'-x' \right)  
\end{align}
Exploiting hypothesis \ref{hyp:positivity_coefficients} and re-establishing the symmetry between $x$ and $y$, the previous expression becomes:
\begin{equation}
    \textbf{G}_{x, t+1} = \frac{1}{2\,\textbf{E}_{p_t}\left[ \gamma_t(x) \right]} \int dx dy\, p_t(x) p_t(y)\, \left| \gamma_t(x) - \gamma_t(y) \right| + \frac{\mathbf{E}_{p_t}\left[F_t\left( x, y\right) \right]}{\textbf{E}_{p_{t}}\left[ \gamma_t(x) \right]  } 
\end{equation}
and we proved the desired result.

\section{Bounds on $\mathbf{E}_{p_t}\left[ F_t(x,y) \right]$ }
\label{app:f_bounds}
Here we want to prove the existence of upper and lower bounds for $\mathbf{E}_{p_t}\left[ F_t(x,y) \right]$, of the form:
\begin{equation}
    \mathbf{E}_{p_t}\left[ F_t(x,y) \right] \leq 2\, \int_0^{\infty}\, dy\, p_t(y)\,P_t\left( x\geq y\right)\, \sigma_t(y) 
\end{equation}
with:
\begin{equation}
    \sigma_t(y) \coloneqq \int_0^{\infty} dy'\,w_t(y'|y)  \,|y' - \gamma_t(y)|
\end{equation}
and:
\begin{equation}
\mathbf{E}_{p_t}\left[ F_t(x,y) \right] \geq \Omega\left( \Gamma_t\right)  \kappa\, \textbf{E}_{p_{t}}\left[ x \right]\, \textbf{P}_{t}\left( x \geq \kappa\, \textbf{E}_{p_{t}}\left[ x \right] \right)^2
\end{equation}
with $\Omega\left( \Gamma_t\right)$ a positive function of $\Gamma_t$ and $\kappa$ an arbitrary constant.

Starting with the proof of the upper bound, and considering without loss of generality $x>y$, we have:
\begin{align}
\label{eq:f_decrescente}
    F_t(x,y) = & \int_{y'>x'} dx'dy'\,w_t(x'|x) w_t(y'|y) \,|x' - y'| \\
    = & \int dy'\,w_t(y'|y) \int_{0}^{y'} dx'\,w_t(x'|x)  \,\left( y' - x'  \right) 
\end{align}
The form of the stochastic growth \eqref{general_stochastic_growth} implies that the probability $\textbf{P}_{w_t(x'|x)}\left( x' \geq \underline{x} \right)$ is a non-decreasing function of $x$ for every $\underline{x}$. Indeed, we can write $w_t(x'|x)$ as:
\begin{align}
    w_t(x'|x) = & \int da\,db\, \rho_t\left(a, b \right)\,\delta\left(x' - a\,x - b \right) = \\
    = & \int \frac{db}{x}\, \rho_t\left(\frac{x' - b}{x}, b \right)
\end{align}
where $\rho_t\left(a, b \right)$ is the joint distribution of the two processes $\hat{A}_t$ and $\hat{B}_t$. The probability $\textbf{P}_{w_t(x'|x)}\left( x' \geq \underline{x} \right)$ can then be written as:
\begin{align}
    \textbf{P}_{w_t(x'|x)}\left( x' \geq \underline{x} \right) & = \int_{\underline{x}}^{\infty} \frac{dx'}{x}\,\int db\, \rho_t\left(\frac{x' - b}{x}, b \right) \\
    & = \int db \, \int_{\frac{\underline{x} - b}{x}}^{\infty} da\, \rho_t\left(a, b \right)
\end{align}
The distribution $\rho_t\left(a, b \right)$ is a non-negative quantity, hence the integral $\int_{\frac{\underline{x} - b}{x}}^{\infty} da\, \rho_t\left(a, b \right)$ is a non-decreasing function of $x$ for all values of $b$.
The fact that the probability $\textbf{P}_{w_t(x'|x)}\left( x' \geq \underline{x} \right)$ is a non-decreasing function of $x$ in turn implies that for any non-decreasing piecewise differentiable function $f(x')$ we have:
\begin{equation}
    x_1\leq x_2 \Longrightarrow \textbf{E}_{w_t(x'|x_1)}\left[ f(x') \right] \leq  \textbf{E}_{w_t(x'|x_2)}\left[ f(x') \right]
\end{equation}
In fact, integrating by part we can write:
\begin{align}
    \textbf{E}_{w_t(x'|x_1)}\left[ f(x') \right] & = \int dx' \, f(x')\, w_t(x'|x) = \\
    & =  f(0) + \int dx' \, \frac{d f(x')}{dx'}\, \int_{x'}^{\infty} dx''\, w_t(x''|x)
\end{align}
The right-hand side of the identity is the integral of a non-decreasing function of $x$ multiplied by a non-negative term, which implies that the integral itself is non-decreasing. 

In equation \eqref{eq:f_decrescente} we can write:
\begin{equation}
    \int_{0}^{y'} dx'\,w_t(x'|x)  \,\left( y' - x'  \right) = \int dx'\,w_t(x'|x)  \, \left( y' - x'  \right) \, \mathbb{1}\left( x' \leq y' \right)
\end{equation}
and the function $\left( y' - x'  \right) \, \mathbb{1}\left( x' \leq y' \right)$ is piecewise differentiable and non-increasing. This implies that the whole integral is a non-increasing function of $x$. Hence from equation \eqref{eq:f_decrescente} we obtain, for any $x\geq y$:
\begin{equation}
    F_t\left( x, y\right) \leq F_t\left( y, y\right)
\end{equation}
From this we have:
\begin{align}
    \mathbf{E}_{p_t}\left[ F_t(x,y) \right] & = \frac{1}{2}\int dx dy\, p_{t}(x) p_{t}(y)\, F_t(x,y)  = \\
    & = \int_{x>y} dx dy\, p_{t}(x) p_{t}(y)\, F_t(x,y) \leq \\
    & \leq \int_{x>y} dx dy\, p_{t}(x) p_{t}(y)\, F_t(y,y) = \\
    & = \int dy\, p_{t}(y) \textbf{P}\left(x\geq y \right)\, F_t(y,y)
\end{align}
Finally, we have:
\begin{align}
    F_t(y,y) & = \int dx'dy'\,w_t(x'|y) w_t(y'|y) \,|x' - y'| \leq \\
    & \leq 2 \int dx'\,w_t(x'|y) \,|x' - \gamma_t(y)| = \\
    & = 2 \, \sigma_t(y)
\end{align}
and the desired result:
\begin{equation}
    \mathbf{E}_{p_t}\left[ F_t(x,y) \right] \leq 2\, \int dy\, p_{t}(y) \textbf{P}\left(x\geq y \right)\, \sigma_t(y)
\end{equation}

We can now pass to the proof of the existence of the lower bound:
\begin{equation}
\mathbf{E}_{p_t}\left[ F_t(x,y) \right] \geq \Omega\left( \Gamma_t\right)  \kappa\, \textbf{E}_{p_{t}}\left[ x \right]\, \textbf{P}_{t}\left( x \geq \kappa\, \textbf{E}_{p_{t}}\left[ x \right] \right)^2
\end{equation}
whenever the ratio $\varrho_t$ tends to zero. 
Considering the trivial identity $ \textbf{E}\left[  \hat{B}_t(x) \right] =  \textbf{E}_{p_t}\left[  \beta_t(x) \right]$, we obtain by Markov's inequality:
\begin{equation}
    \textbf{P}\left(\hat{B}_t(x) \geq x\, \frac{\textbf{E}_{p_t}\left[  \beta_t(x) \right] }{\textbf{E}_{p_t}\left[  \alpha_t(x)\, x \right]} \right) \leq \frac{ \textbf{E}_{p_t}\left[  \alpha_t(x)\, x \right] }{ x } 
\end{equation}
From this, taking the limit $\frac{x}{\textbf{E}_{p_t}\left[  \alpha_t(x)\, x \right]}\to \infty$ we obtain, for any finite $\varrho_t$,  the condition:
\begin{equation}
    \lim_{x\to \infty} \textbf{P}\left(\frac{\hat{B}_t(x)}{x} \geq \varrho_t  \right) = 0
\end{equation}
This in turn implies that for large values of wealth and $\varrho_t \to 0$ the growth process becomes essentially multiplicative. In terms of the conditional probability $w_t(x'|x)$, this can be expressed as:
\begin{equation}
\label{eq:multiplicative_limit}
    \lim_{x\to \infty} x\,w_t(a\,x |x) = \rho_t\left( a \right)
\end{equation}
We want to derive this result to obtain some useful regularity conditions on $F_t(x,y)$ and $p_t(x)$. 
Starting with $F_t(x,y)$, without loss of generality we consider the region $x\geq y$:
\begin{equation}
    F_t(x,y) = \int_{y'>x'} dx'dy'\,w_t(x'|x) w_t(y'|y) \,|x' - y'| 
\end{equation}
We can divide the region of integration in two sub-regions characterized be the value of the logarithmic derivative of $w_t(y'|y)$:
\begin{equation}
    F_t(x,y) = \left( I_1 + I_2 \right)
\end{equation}
with:
\begin{equation}
    I_1 = \int_{y'>x'\, ,\, \left| \frac{\partial \log w_t(y'|y)}{\partial \log y}\right| \leq L} dx' dy'\, w_t(x'|x)\, w_t(y'|y) \,|x' - y'|
\end{equation}
\begin{equation}
    I_2 = \int_{y'>x'\, ,\, \left| \frac{\partial \log w_t(y'|y)}{\partial \log y}\right| > L} dx' dy'\, w_t(x'|x)\, w_t(y'|y)  \,|x' - y'|
\end{equation}
For the second integral, we can write:
\begin{equation}
    I_2 \leq \int_{x'\leq y'\leq y \, , \, \left| \frac{\partial \log w_t(y'|y)}{\partial \log y}\right| > L} dy'\,  w_t(y'|y)  \, y' +    \int_{y' >  \max{(x',y)} \, , \, \left| \frac{\partial \log w_t(y'|y)}{\partial \log y}\right| > L} dy'\,  w_t(y'|y)\, y'
\end{equation}
For any finite $y$, for $\left| \frac{\partial \log w_t(y'|y)}{\partial \log y}\right|$ to be larger than $L$ for large enough $L$ we need $y'\to \infty$ or $y'\to 0$ (we assume the logarithmic derivatives to be continuous, and as such bounded in any finite region). 
This implies the two integrals can be written as:
\begin{equation}
    I_2 \leq  \int_{y' < m_L(y)\,y} dy'\,  w_t(y'|y)  \, y' +    \int_{y'> M_L(y)\,y } dy'\,  w_t(y'|y)\, y'
\end{equation}
The two integrals can than be made small at will. 
On the other hand, by equation \eqref{eq:multiplicative_limit}, for large enough values of $y$ we can describe $w_t(y'|y)$ with a multiplicative process: 
\begin{equation}
    w_t(y'|y) \approx \frac{1}{y}\,\rho_t\left( \frac{y'}{y} \right)
\end{equation}
Hence a divergence of $\left| \frac{\partial \log w_t(y'|y)}{\partial \log y}\right|$ for large $y$ implies the divergence of $\left| \frac{\partial \log \rho_t\left(a \right)}{\partial \log a}\right|$ with $a=\frac{y'}{y}$, and as a consequence a divergence of the ratio $\frac{y'}{y}$. 
The condition $\left| \frac{\partial \log \rho_t\left(a \right)}{\partial \log a}\right| > L$ is equivalent for large $L$ to:
\begin{equation}
    y' < m_{L,\rho} \, y \quad \text{or} \quad y' > M_{L,\rho} \, y
\end{equation}
for some constants $m_{L,\rho}$ and $M_{L,\rho}$. 
From this we can see that the functions $m_L(y)$ and $M_L(y)$ have finite limits when $y\to \infty$. 
In addition we are interested in a region limited to $x \geq \kappa\,\textbf{E}_{p_t}\left[ x'\right]$, hence potential divergences for $y \to 0$ are inconsequential. 
Being the two functions continuous, they can be bounded for any $y$.
The convergence of the two integrals in $I_2$ to zero can be thus made uniform, and we can write:
\begin{equation}
    \lim_{L \to \infty} \frac{I_2}{y} = 0 
\end{equation}
For $I_1$, whenever $\left| \log \left(\frac{x}{y}\right) \right| \leq \delta$, we can write:\footnote{We have that $\frac{w_t(y'|y)}{w_t\left(y'| x \right)} \geq \exp - L \left| \log \left(\frac{x}{y}\right) \right| \geq \exp - L \delta$; the last expression is greater than $(1 - \epsilon)$ for small enough $\delta$.}
\begin{align}
    I_1 & \geq (1 - \epsilon) \int_{y'>x'\, ,\, \left| \frac{\partial \log w_t\left( y' |x \right)}{\partial \log x}\right| \leq L} dx' dy'\, w_t(x'|x)\, w_t\left( y'|x \right) \,|x' - y'| \geq \\
    & \geq  (1 - \epsilon) \left[ F_t(x,x) - \epsilon'\,x \right]
\end{align}
where we exploited the smallness of $I_2$ for large $L$. As we will show in the following, $F_t(x,x)$ is bounded by:
\begin{equation}
    F_t(x,x) \geq \Gamma_t\, x
\end{equation}
for some constant $\Gamma_t$, hence for small enough $\epsilon'$ we obtain:
\begin{equation}
    \forall \, \epsilon>0 \; \exists \; \delta>0 \, : \, \left| \log \left(\frac{x}{y}\right) \right| \leq \delta \Longrightarrow F(x,y) \geq (1 - \epsilon) \, F(x,x) 
\end{equation}
In a similar way, we can write for the probability $p_t(x)$:
\begin{align}
    p_t(x') & = \int dx\, p_{t-1}(x)\,w_t(x'| x) = \\
    & = \int_{\left| \frac{\partial \log w_t\left(x'|x \right)}{\partial \log x'}\right| \leq L} dx\, p_{t-1}(x)\,w_t(x'| x) + \int_{\left| \frac{\partial \log w_t\left(x'|x \right)}{\partial \log x'}\right| > L} dx\, p_{t-1}(x)\,w_t(x'| x)
\end{align}
We assume for the process $\hat{B}_t(x)$ to have a non-trivial distribution in the limit $x\to 0$, such that to have $\left| \frac{\partial \log w_t\left(x'|x \right)}{\partial \log x'}\right| > L$ for large enough $L$ and fixed $x'$, we must have $x'< m_L(x)$ for some function $m_L(x)$ going zero with $L$.
Inverting this function we obtain $x> m_L^{-1}(x')$, with:
\begin{equation}
    \lim_{L\to \infty} m_L^{-1}(x') = \infty
\end{equation}
Finally we can exploit the convergence to zero of the distribution $w_t\left(x'|x \right)$ for $x\to 0$ and $x\to \infty$ for any fixed $x'$, to obtain that the second integral can be made small at will:
\begin{align}
    & \int_{\left| \frac{\partial \log w_t\left(x'|x \right)}{\partial \log x'}\right| > L} dx\, p_{t-1}(x)\,w_t(x'| x) \leq \\
    & \leq \int_{x>m_L^{-1}(x')} dx\, p_{t-1}(x)\,w_t(x'| x) \approx \\
    & \approx \frac{1}{m_{L, \rho}'\, x'}\int_{x> m_{L, \rho}'\, x'} dx\, p_{t-1}(x)\,\rho_t\left(\frac{x'}{x} \right) \leq \\
    & \leq  \frac{\epsilon_{L,\rho}}{m_{L, \rho}'\, x'} \int_{\left| \frac{\partial \log w_t\left(y \to x \right)}{\partial \log x}\right| > L} dy\, p_{t-1}(y) \leq \\
    & \leq \frac{\epsilon_{L,\rho}}{m_{L, \rho}'\, x'} 
\end{align}
We limit the proof to the region $x' \geq \kappa\, \textbf{E}_{p_t}\left[ x \right]$, such that the last term can be bounded with an arbitrarily small constant $\epsilon$, and we obtain:
\begin{equation}
    p_t(x) = \int_{\left| \frac{\partial \log w_t\left(y \to x \right)}{\partial \log x}\right| \leq L} dy\, p_{t-1}(y)\,w_t(y\to x) + \epsilon
\end{equation}
From this identity, whenever $\left| \log \left(\frac{x}{y}\right) \right|<\delta$, we have:
\begin{equation}
    p_t(x) \geq p_t(y)\, \left( 1- \epsilon'\right)
\end{equation}
for arbitrarily small $\epsilon'$ (and $p_t(x)\gg \epsilon$).

Finally, hypothesis \ref{hyp:positive_dispersion} allows us to put a lower bound on the value of $F_t(x,x)$. Exploiting properties of the absolute value we can write:
\begin{equation}
    F_t(x,x) \geq   \int dx' \, w_t(x' |x)\,  |x' - \gamma_t(x)| = \sigma_t(x)
\end{equation}
As already stated, we are interested in the region $x\geq \kappa \, \textbf{E}\left[  x' \right]$. Hypothesis \ref{hyp:positive_dispersion} implies the existence of a value of $\kappa$ such that:
\begin{equation}
    \sigma_t(x) \geq \Gamma_t\, x
\end{equation}
for every $x$ greater than $\kappa \, \textbf{E}\left[  x' \right]$. Selecting this as the lower bound of our region of interest, we can establish the desired bound:
\begin{equation}
    F_t(x,x) \geq \Gamma_t\, x
\end{equation}

We now exploit the constraints found in the rest of the section to prove inequality \eqref{eq:main_inequality}:
\begin{equation}
    \mathbf{E}_{p_t}\left[ F_t(x,y) \right]  \geq \Omega\left( \Gamma_t \right) \,\kappa\, \textbf{E}_{p_{t}}\left[ x \right]\, P_{t}\left( x \geq \kappa\, \textbf{E}_{p_{t}}\left[ x \right] \right)^2
\end{equation}
with the expectation value defined as:
\begin{equation}
    \mathbf{E}_{p_t}\left[ F_t(x,y) \right] \coloneqq \int_{x > y} dx dy\, p_{t}(x) p_{t}(y)\, F_t(x,y) 
\end{equation}
The integrand on the right hand side is always non-negative, such that restricting the integral to any proper sub-region we obtain a lower value. For any region $R_{\delta}$:
\begin{equation}
    \mathbf{E}_{p_t}\left[ F_t(x,y) \right] \geq P_{t}\left((x,y) \in R_{\delta} \right) \mathbf{E}_{p_t}\left[ F_t(x,y) \big| (x,y) \in R_{\delta}\, \right]   
\end{equation}
Here we select the region of integration:
\begin{equation}
    R_{\delta} =  \left\{(x,y) : \left| x-y\right| < \delta x\right\} \cap \left\{ (x,y) \in \left(\kappa\, \textbf{E}_{p_{t}}\left[ x \right], \infty \right)^2 \right\}
\end{equation}
which implies:
\begin{equation}
    \mathbf{E}_{p_t}\left[ F_t(x,y) \right] \geq P_{t}\left( x \geq \kappa\, \textbf{E}_{p_{t}}\left[ x \right] \right)^2 \mathbf{E}_{p_t}\left[ F_t(x,y) \big| (x,y) \in R_{\delta}\, \right], 
\end{equation}
where we defined:
\begin{equation}
    P_{t}\left( x \geq \kappa\, \textbf{E}_{p_{t}}\left[ x \right] \right) \coloneqq \int_{\kappa \textbf{E}_{p_{t}}\left[ x \right]}^{\infty} dx \, p_t(x),
\end{equation}
\begin{equation}
\label{stripe_expectation}
    \mathbf{E}_{p_t}\left[ F_t(x,y) \big| (x,y) \in R_{\delta}\, \right] \coloneqq \int_{(x,y) \in R_{\delta}} dx dy\, \tilde{p}_{t}(x) \tilde{p}_{t}(y)\, F_t(x,y)
\end{equation}
and the rescaled probability density:
\begin{equation}
    \tilde{p}_{t}(x) \coloneqq \frac{p_t(x)}{ P_{t}\left( x \geq \kappa\, \textbf{E}_{p_{t}}\left[ x \right] \right)  } \, , \quad \int_{\kappa \textbf{E}_{p_{t}}\left[ x \right]}^{\infty} dx \, \tilde{p}_t(x) = 1
\end{equation}
We are thus left to prove:
\begin{equation}
    \mathbf{E}_{p_t}\left[ F_t(x,y) \big| (x,y) \in R_{\delta}\, \right]  \geq  \kappa\, \textbf{E}_{p_{t}}\left[ x \right]\, \Omega\left( \Gamma_t \right)
\end{equation}
for some positive $\Omega\left( \Gamma_t \right)$.
With the bounds found in the last section, we can write:
\begin{align}
      \mathbf{E}_{p_t}\left[ F_t(x,y) \big| (x,y) \in R_{\delta}\, \right] & \geq  \left( 1 - \epsilon  \right) \int_{ (x,y) \in R_{\delta}}dx dy\, \tilde{p}_t(x)^2\,   F_t(x,x) \\
      & = 2\, \delta\, \left( 1 - \epsilon \right) \int_{\kappa \textbf{E}_{p_{t}}\left[ x \right]}^{\infty} dx \, \tilde{p}_t(x)^2\,x\,   F_t(x,x) + o(\delta) \\
      & \geq 2\, \delta\, \left( 1 - \epsilon \right)\, \Gamma_t \int_{\kappa \textbf{E}_{p_{t}}\left[ x \right]}^{\infty} dx \, \tilde{p}_t(x)^2\,x^2 + o(\delta) 
\end{align}
Extremizing the convex functional: 
\begin{equation}
\label{funzionale}
Y_{a}\left[p \right] \coloneqq \int_{a}^{\infty} dx\, p(x)^2\, x^2
\end{equation}
over the space of probability distributions $p(x)$ defined on the interval $\left( a, \infty \right)$, we find:
\begin{equation}
    Y_{a}\left[p \right] \geq a
\end{equation}
with extremal point coinciding with the distribution $p(x) = \frac{a}{x^2}$. Applying this result to our case, and neglecting higher order terms in $\delta$ and terms proportional to $\epsilon$, we obtain the desired result:
\begin{equation}
    \mathbf{E}_{p_t}\left[ F_t(x,y) \big| (x,y) \in R_{\delta}\, \right] \geq \kappa\, \textbf{E}_{p_{t}}\left[ x \right]\, \Omega\left( \Gamma_t \right),
\end{equation}
with $\Omega\left( \Gamma_t \right) = 2\, \delta\, \Gamma_t$.
This in turn implies:
\begin{equation}
    \mathbf{E}_{p_t}\left[ F_t(x,y) \right] \geq \Omega\left( \Gamma_t \right) \, \kappa \,\textbf{E}_{p_{t}}\left[ x \right]\,  P_{t}\left( x \geq \kappa\, \textbf{E}_{p_{t}}\left[ x \right] \right)^2 
\end{equation}

\section{Bound on $\textbf{G}_{\alpha x, t}$}
\label{app:alpha_growth}
In this appendix we want to prove the inequality:
\begin{equation}
    \textbf{G}_{\alpha x, t} \geq \textbf{G}_{x,t} 
\end{equation}
for any non-decreasing function $\alpha_t(x)$. The index $\textbf{G}_{\alpha x, t}$ is defined as:
\begin{equation}
    \textbf{G}_{\alpha x, t} \coloneqq \frac{1}{2\, \textbf{E}_{p_{t}}\left[ \alpha_t(x)\,x \right]   } \int dx dy\, p_{t}(x) p_{t}(y) \, \left| \alpha_t(x)\,x - \alpha_t(y)\,y \right|
\end{equation}
but it can be equivalently defined as twice the area between the Lorenz curve of the function $\alpha_t(x)\,x$ and the line of perfect equality. 
The Lorenz curves of the functions $x$ and $\alpha_t(x)\,x$ are defined as:
\begin{equation}
    L_x(z) = \frac{  1 }{\mathbf{E}_{p_t}\left[ x' \right]} \int_0^z dx' \,p_t(x')\,x'
\end{equation}
\begin{equation}
    L_{\alpha x}(z) = \frac{ 1  }{\mathbf{E}_{p_t}\left[ \alpha_t(x')\,x' \right]} \int_0^z dx'\,p_t(x')\,\alpha_t(x')\,x'
\end{equation}
To prove inequality \eqref{eq:financial_gini_bound} it is sufficient to prove that the the curve $L_{\alpha x}(z)$ is always dominated by $L_x(z)$, as in this case the area between $L_{\alpha x}(z)$ and the line of perfect equality will always be greater than the one between $L_x(z)$ and the same line. 
To this end, we note that the condition 
\begin{equation}
    \frac{d L_{\alpha x}(z)}{d z} \leq \frac{d L_x(z)}{d z}    
\end{equation} 
is equivalent to:
\begin{equation}
    \alpha_t(x) \leq E_{\tilde{p}_t} \left[ \alpha_t(x)\right],
\end{equation}
where we introduced the rescaled probability:
\begin{equation}
    \tilde{p}_t(x) \coloneqq \frac{p_t(x)\,x}{E_{\tilde{p}_t} \left[ x'\right]}.
\end{equation}
If $\alpha_t(x)$ is a non-decreasing function the last inequality is satisfied for $x=0$. In addition we have $L_{\alpha x}(0) = L_{x}(0) = 0$, which implies $L_{\alpha x}(z) \leq L_x(z)$ for small enough values of $z$.
Conversely, the inequality:
\begin{equation}
    \frac{d L_{\alpha x}(z)}{d z} > \frac{d L_x(z)}{d z}
\end{equation}
admits as solution the whole connected interval $z \in \left( \alpha_t^{-1}\left(  E_{\tilde{p}_t} \left[ \alpha_t(x)\right] \right), \, \infty \right)$, such that the two curves cross for $z>0$ only in one point, corresponding by definition to $z=\infty$. Hence the two curves, starting in the order $L_{\alpha x}(z) \leq L_x(z)$ in a neighborhood of zero, respect the same order for each $z < \infty$. 
We proved than that if $\alpha_t(x)$ is a non-decreasing function of $x$, the Lorenz curve $L_{\alpha x}(z)$ is dominated by $L_x(z)$, which in turn implies the inequality:
\begin{equation}
    \textbf{G}_{\alpha x, t} \geq \textbf{G}_{x,t}.
\end{equation}

\section{Properties of $\alpha_t(x)$ and $\beta_t(x)$}
\label{app:positive_beta}
In this appendix we discuss the validity of hypotheses \ref{hyp:positivity_coefficients}, \ref{hyp:positive_dispersion} for all the models we take into consideration. In general the easiest and most general way to check the validity of our hypotheses for a given model is to analyze a given realization of its dynamics, and evaluate the two average functions $\alpha_t(x)$ and $\beta_t(x)$ from the available data. In some cases of interest however analytical results are available, and the validity of our hypotheses can be proved without the need of a particular realization of the dynamics.\footnote{For models with infinite memory the validity of hypothesis \ref{hyp:positivity_coefficients} is conditional to the initial distribution of wealth and labor income. We assume in this case a null wealth initial value with probability $1$.} 
A series of useful results for this purpose is proved in \cite{efron1965increasing}, \cite{lehmann1966some}, and more recently \cite{pellerey2022stochastic}.
Here we will exploit the simplest of these results, Efron's theorem, proved in \cite{efron1965increasing}.
Efron's theorem shows that, given an arbitrary number $n$ of independent random variables $X_i$ having log-concave probability densities,\footnote{A non-negative function is said to be log-concave if its logarithm is a concave function.} defining their sum:
\begin{equation}
    S_n = \sum_{i=1}^n X_i,
\end{equation}
for any real measurable function $\phi$ non-decreasing in each of its arguments we have that the expectation value:
\begin{equation}
    \textbf{E}\left[\left. \phi\left(X_1, X_2, ..., X_n \right)  \right| S_n = s  \right]    
\end{equation}
is a non-decreasing function of $s$.
We will see in the following subsections that for model \textbf{M1} and \textbf{M2} we can write:
\begin{equation}
    \log \hat{x}_t = \log \hat{y} + \log \hat{Z}_t
\end{equation}
for some random variable $\hat{Z}_t$ independent of $\hat{y}$, with both having a log-concave probability density.  
For model \textbf{M3} the two random variables $\hat{y}$ and $\hat{Z}_t$ cannot be considered independent, and the novel result proved in \cite{pellerey2022stochastic} must be used to prove the monotonicity of $\beta_t(x)$. 
It is however much easier to generate a sample from the model and check the validity of the hypotheses on the resulting distribution.
We performed the check in a wide range of parameters, finding the hypothesis of a monotonous $\beta_t(x)$ always confirmed. 
The monotonicity of $\alpha_t(x)$ and hypothesis \ref{hyp:positive_dispersion}, on the other hand, are immediate to check analytically.

\subsection{Model \textbf{M1}}
The evolution of this model is given by:
\begin{equation}
    \hat{x}_{t+1} = e^{\hat{r}_t}\, \hat{x}_t + \hat{y}_t 
\end{equation}
\begin{equation}
    \hat{y}_t = e^{\dot{\mu}_y}\, \hat{y}_{t-1}
\end{equation}
where $\hat{x}_0 = 0$ with probability $1$, and $\hat{y}_t$ has a log-normal initial distribution. 
For any finite $n$ we obtain, iterating the evolution equation:
\begin{equation}
    \hat{x}_{t+n} = \hat{y}_{t+n-1}\, \hat{Z}_{t,n} 
\end{equation}
\begin{equation}
    \hat{Z}_{t,n} = \left[ 1 + \sum_{k=1}^{n-1} \exp{\sum_{j_k=1}^{k}\tilde{r}_{t+n-j_{k}}}\right]
\end{equation}
where we used the shorthand:
\begin{equation}
    \tilde{r}_{t} = \hat{r}_t  - \dot{\mu}_y
\end{equation}
The quantity $\hat{Z}_{t,n}$ is independent of $y_{t}$, and we can write:
\begin{equation}
    \log \hat{x}_{t+n} = \log \hat{y}_{t+n-1} + \log \hat{Z}_{t,n}.
\end{equation}
We see that the logarithm of $\hat{x}_t$ is the sum of two independent random variables, one of which ($\log \hat{y}_{t+n-1}$) distributed normally. It is easy to check numerically the log-concavity of the probability density of $\log \hat{Z}_{t,n}$, hence we can apply Efron's theorem to conclude that for every non-decreasing function $\phi$ of $\log \hat{y}_{t+n-1}$ and $\log \hat{Z}_{t,n}$ we have:
\begin{equation}
    \frac{d}{d x} \textbf{E}\left[\left. \phi\left(\log \hat{y}_{t+n-1}, \, \log \hat{X}_{t,n}\right) \right|x_{t+n} = x  \right] \geq 0
\end{equation}
This of course implies:
\begin{equation}
    \frac{d}{d x} \beta_t(x) = \frac{d}{d x} \textbf{E}\left[\left. e^{\log \hat{y}_{t}} \right| x_{t} = x  \right] \geq 0
\end{equation}
In addition, we have $\alpha_t(x) = \textbf{E}\left[ e^{\hat{r}_t} \right]$, such that:
\begin{equation}
    \frac{d}{d x} \alpha_t(x) =  0.
\end{equation}
Finally, hypothesis \ref{hyp:positive_dispersion} is trivially satisfied for $\sigma_r >0$.

\subsection{Model \textbf{M2}}
Model \textbf{M2} can be written as:
\begin{equation}
    \hat{x}_{t+1} = s_t\left( \frac{\hat{x}_t}{\hat{y}_{t-1}}\right)e^{\hat{r}_t}\, \hat{x}_t + \hat{y}_t 
\end{equation}
\begin{equation}
    \hat{y}_{t+1} = \hat{y}_{t}
\end{equation}
where again $\hat{r}_t$ has normal distribution and $\hat{y}_t$ has a log-normal initial distribution.
Equation \eqref{eq:log_contraction} implies that the system has a finite memory, meaning that for each agent any wealth perturbations away from equilibrium is consumed in a finite time. In particular the initial value of wealth is irrelevant after a finite time, and for $n$ large enough the dynamics reduces with probability arbitrarily close to one to:
\begin{equation}
    \hat{x}_{t+n} = \hat{y}_{t+n-1} \, \frac{ \hat{x}_{t+n} }{\hat{y}_{t+n-1}}
\end{equation}
\begin{equation}
\label{eq:evo_decomposition}
    \frac{ \hat{x}_{t+n} }{\hat{y}_{t+n-1}} = \left[ 1 + \sum_{k=1}^{n-1} \exp{\sum_{j_k=1}^{k}\tilde{r}_{t+n-j_{k}}}\right]
\end{equation}
where we used the shorthand:
\begin{equation}
    \tilde{r}_{t} = \hat{r}_t  + \log s_t\left(\frac{\hat{x}_{t}}{\hat{y}_{t-1}}\right)
\end{equation}
We showed in section \ref{sec:m2} that the evolution of the ratio $\hat{x}_{t+n}/\hat{y}_{t+n-1}$ is independent of the value of $\hat{y}_{t+n-1}$, such that the random variable:
\begin{equation}
    Z_{t,n} \coloneqq  \frac{ \hat{x}_{t+n} }{\hat{y}_{t+n-1}} = \left[ 1 + \sum_{k=1}^{n-1} \exp{\sum_{j_k=1}^{k}\tilde{r}_{t+n-j_{k}}}\right]
\end{equation}
is itself independent of $\hat{y}_{t+n-1}$.
We can write the stochastic evolution as:
\begin{equation}
    \log \hat{x}_{t+n} = \log \hat{y}_{t+n-1} + \log \hat{Z}_{t,n}
\end{equation}
and again apply Efron's theorem to obtain that any expectation value:
\begin{equation}
    \textbf{E}\left[\left. \phi\left(\log \hat{y}_{t+n-1}, \, \log \hat{Z}_{t,n}\right) \right|x_{t+n} = x  \right]
\end{equation}
is non-decreasing in $x$ for any measurable function $\phi$ non-decreasing in each of its arguments. 
This implies, in particular:
\begin{equation}
    \frac{d}{d x }\textbf{E}\left[\left. e^{\log \hat{y}_{t}} \right| x_{t} = x  \right] \geq 0
\end{equation}
\begin{equation}
    \frac{d}{d x }\textbf{E}\left[\left. s_t\left( \frac{ \hat{x}_{t} }{\hat{y}_{t-1}} \right) \right| x_{t} = x  \right] \geq 0
\end{equation}
Hence we proved the validity of hypothesis \ref{hyp:positivity_coefficients}.
Regarding hypothesis \ref{hyp:positive_dispersion}, for $x \to \infty$ we have:
\begin{equation}
    \lim_{x\to \infty} \textbf{E}\left[ \left. e^{\hat{r}_t}\, s_t\left(\frac{ \hat{x}_t }{ \hat{y}_{t - 1} }\right) \right| \hat{x}_t=x \right] = \eta\, \textbf{E}\left[ e^{\hat{r}_t}\right] > 0.
\end{equation}
Finally, we want to prove that equation \eqref{eq:contraction} implies the existence of an asymptotic average wealth. 
To this aim, we can write equation \eqref{eq:evo_decomposition} as:
\begin{equation}
    \hat{x}_{t}  = \hat{y}_{t-1}\,\left[ 1 + \sum_{k=1}^{t-1} \prod_{j=t-k}^{t-1} s_j\left( \frac{\hat{x}_j}{\hat{y}_{t-1}}\right)e^{\hat{r}_j} \right]
\end{equation}
Taking the average of the last expression we obtain:
\begin{equation}
\label{eq:geometric_series}
    \textbf{E}\left[ \hat{x}_t \right] \leq \textbf{E}\left[ \hat{y}_{t-1} \right] \, \sum_{k=0}^{t-1} \psi^k 
\end{equation}
with $\psi = \eta\, \textbf{E}\left[ e^{\hat{r}_t} \right]$.
If equation \eqref{eq:contraction} is assumed to be valid, we have $\psi < 1$. 
Hence the geometric series in equation \eqref{eq:geometric_series} has a finite limit for $t \to \infty$, implying a finite asymptotic average wealth.


\section{Properties of the CRRA consumption function}
\label{app:consumption}
In this appendix we prove two properties of the solution of the dynamic optimization problem, derived from the CRRA utility function \eqref{eq:utility}:
\begin{equation}
    u(c_t) = \frac{c_t^{1-\gamma}}{1-\gamma} \quad , \quad \gamma >0
\end{equation}
The optimal consumption strategy with this utility function is given by the solution of the Bellman equation:
\begin{equation}
    V(x_t, y_{t-1}) = \max_{c_t \leq x_t} \left\{ u(c_t) + \beta\,\textbf{E}\left[ V\left( e^{\hat{r}_t}\,\left( x_t - c_t \right) + \hat{z}_t\, y_{t-1} , \,\hat{z}_t \, y_{t-1} \right) \right] \right\}
\end{equation}
where we assumed the labor income $\hat{y}_t$ to be a homogeneous function of $y_{t-1}$:
\begin{equation}
    \hat{y}_t = \hat{z}_t\, y_{t-1} 
\end{equation}
for some positive random variable $\hat{z}_t$ independent of $y_{t-1}$ and $x_t$, and the returns on wealth $\hat{r}_t$ follow a normal distribution: $\hat{r}_t \sim N\left(\mu_r , \sigma_r\right)$.

The first result we want to prove is a homogeneity property of the value function\footnote{A similar derivation is given in \cite{toda2017discrete} for a purely multiplicative dynamics.} $V(x_t, y_{t-1})$:
\begin{equation}
    V(x_t, y_{t-1}) = \tilde{V}\left( \frac{x_t}{y_{t-1}}\right) \, x_t^{1-\gamma}
\end{equation}
To prove this result, we note that starting with wealth and labor income values $\lambda x_t$ and $\lambda y_{t-1}$, it is possible for the optimizing agent to choose a consumption policy $\lambda c_{t'}$ for all $t'\geq t$, where $c_{t'}$ is the optimal consumption policy starting with values $x_t$ and $y_{t-1}$.
In addition, defining the auxiliary function:
\begin{equation}
    \underline{V}(c_t) =  \sum_{T=t}^{\infty} \beta^{T}\,\textbf{E}\left[ u(c_T) \right]
\end{equation}
we have:
\begin{equation}
    \underline{V}(\lambda c_t) = \lambda^{1-\gamma}\, \underline{V}(c_t)
\end{equation}
and by the very definition of the value function $V(x_t, y_{t-1})$:
\begin{equation}
    V(\lambda x_t, \lambda y_{t-1}) \geq \underline{V}(\lambda c_t) 
\end{equation}
If $c_t$ is the optimal consumption policy with respect to the value function $V(x_t, y_{t-1})$, we finally have:
\begin{equation}
    V(x_t, y_{t-1}) = \underline{V}(c_t)
\end{equation}
and combining all these facts we obtain:
\begin{equation}
\label{eq:push_down}
    V(\lambda x_t, \lambda y_{t-1}) \geq \lambda^{1-\gamma}\, V(x_t, y_{t-1})
\end{equation}
Defining the new variables $\lambda' = \lambda^{-1}$, $x'_t = \lambda x_t$, $y'_{t-1} = \lambda y_{t-1}$, the same inequality can be written as:
\begin{equation}
    \left(\lambda'\right)^{1 - \gamma}\, V(x'_t, y'_{t-1}) \geq  V(\lambda' x'_t, \lambda' y'_{t-1})
\end{equation}
and given that the last inequality is valid for any (non-negative) value of $\lambda'$, $x'_t$ and $y'_{t-1}$, we can write it as:
\begin{equation}
    \lambda^{1-\gamma}\, V(x_t, y_{t-1}) \geq  V(\lambda x_t, \lambda y_{t-1})
\end{equation}
This, together with \eqref{eq:push_down}, forces us to conclude:
\begin{equation}
    V(\lambda x_t, \lambda y_{t-1}) = \lambda^{1-\gamma}\, V(x_t, y_{t-1})
\end{equation}
Finally, choosing $\lambda = x_t^{-1}$ and defining the new function:
\begin{equation}
    \tilde{V}\left( \frac{x_t}{y_{t-1}}\right) \coloneqq V\left(1\, ,\, \frac{y_{t-1}}{x_t} \right)
\end{equation}
we have the desired identity:
\begin{equation}
    V(x_t, y_{t-1}) = \tilde{V}\left( \frac{x_t}{y_{t-1}}\right)\, x_t^{1-\gamma}.
\end{equation}
The second result we want to prove is the asymptotic property of the saving function:
\begin{equation}
    \lim_{x_t \to \infty} s_t\left( \frac{x_t}{y_{t-1}} \right)= \eta 
\end{equation}
for some constant $\eta > 0$ (an alternative derivation is given in \cite{benhabib2015wealth}), where:
\begin{equation}
    s_t\left( \frac{x_t}{y_{t-1}} \right) = \left(  1 - \frac{ c_t\left(x_t, y_{t-1} \right) }{x_t}  \right).
\end{equation}
To prove the result we note that the re-scaled value function $\tilde{V}\left( \frac{x_t}{y_{t-1}}\right)$ has the finite limit:
\begin{equation}
    \lim_{x_t \to \infty} \tilde{V}\left( \frac{x_t}{y_{t-1}}\right) = V\left(e^{\hat{r}_t}\,\left( 1 - \tilde{c}_t \right), 0 \right)
\end{equation}
where $\tilde{c}_t$ is the re-scaled optimal consumption policy: $\tilde{c}_t = c_t / x_t$.
This implies that, in this limit, the re-scaled optimal policy $\tilde{c}_t$ cannot depend on either $x_t$ or $y_{t-1}$. In other words, we obtain:
\begin{equation}
    c_t = \zeta\, x_t
\end{equation}
for some constant $\zeta$. 
In addition, solving the Bellman equation (see \cite{benhabib2015wealth, levhari1969optimal}), we find the values:
\begin{equation}
    \zeta = 1 - \left( \beta\, \textbf{E}\left[ e^{(1-\gamma)\, \hat{r}_t}\right] \right)^{\frac{1}{\gamma}} \quad , \quad \eta = \left( \beta\, \textbf{E}\left[ e^{(1-\gamma)\, \hat{r}_t}\right] \right)^{\frac{1}{\gamma}}.
\end{equation}

\section{Divergence of the coefficient of variation}
\label{sec:cv_divergence}
In this section, to show the generality of our derivation, we prove results similar to the ones found in section \ref{sec:gini_evolution} for another widely used inequality index, the coefficient of variation. 
The coefficient of variation is defined as the ratio between the standard deviation and the average of a distribution:
\begin{equation}
    \textbf{CV}_{x,t}^2 \coloneqq \frac{\textbf{Var}_{p_t}\left[ x \right]}{\textbf{E}_{p_t}\left[ x \right]^2}
\end{equation}
With a variance decomposition we can write for the evolution of the coefficient:
\begin{equation}
\label{eq:cv_evolution}
    \textbf{CV}_{x, t+1}^2 = \textbf{CV}_{\gamma, t}^2 + \textbf{C}_{\Gamma, t}^2
\end{equation}
where $\textbf{CV}_{\gamma, t}^2$ is the coefficient of variation of the average evolution:
\begin{equation}
    \textbf{CV}_{\gamma, t}^2 \coloneqq \frac{\textbf{Var}_{p_t}\left[ \gamma_t(x) \right]}{\textbf{E}_{p_t}\left[ \gamma_t(x) \right]^2}
\end{equation}
and $\textbf{C}_{\Gamma, t}^2$ a noise term generated by the dispersion of the evolution around its mean (we denote again as $w_t\left( x' | x\right)$ the conditional probability of having $x_{t+1}=x'$ given $x_t=x$):
\begin{equation}
    \textbf{C}_{\Gamma, t}^2 \coloneqq \frac{\textbf{E}_{p_t}\left[ \textbf{Var}_{w_t}\left[ \left. x' \right| x \right] \right]}{\textbf{E}_{p_t}\left[ \gamma_t(x) \right]^2}
\end{equation}
Similarly to what we did for the Gini index, we can express $\textbf{CV}_{\gamma, t}$ in terms of the two functions $\alpha_t(x)$ and $\beta_t(x)$:
\begin{align}
\label{eq:convex_cv}
    \textbf{CV}_{\gamma, t}^2 = \left( \frac{ 1 }{1 + \varrho_t} \right)^2  & \textbf{CV}_{\alpha x, t}^2 + \left( \frac{\varrho_t}{1 + \varrho_t} \right)^2 \textbf{CV}_{\beta, t}^2 + \frac{2\,\textbf{Cov}_{p_t}\left[ \alpha_t(x)\,x, \,\beta_t(x) \right] }{ \textbf{E}_{p_t}\left[ \gamma_t(x) \right]^2  } \geq \\
    \geq  & \left( \frac{ 1 }{1 + \varrho_t} \right)^2 \textbf{CV}_{\alpha x, t}^2 + \left( \frac{\varrho_t}{1 + \varrho_t} \right)^2 \textbf{CV}_{\beta, t}^2
\end{align}
where for the last inequality we used hypothesis \ref{hyp:positivity_coefficients} and the fact that the covariance of non-decreasing functions is always greater than or equal to zero. 
We used the obvious definitions:
\begin{equation}
    \textbf{CV}_{\alpha x, t}^2 \coloneqq \frac{\textbf{Var}_{p_t}\left[ \alpha_t(x)\,x \right]}{\textbf{E}_{p_t}\left[ \alpha_t(x)\,x \right]^2}
\end{equation}
\begin{equation}
    \textbf{CV}_{\beta, t}^2 \coloneqq \frac{\textbf{Var}_{p_t}\left[ \beta_t(x) \right]}{\textbf{E}_{p_t}\left[ \beta_t(x) \right]^2}
\end{equation}
and, as in section \ref{sec:gini_evolution}:
\begin{equation}
    \varrho_t = \frac{\textbf{E}_{p_t}\left[ \beta_t(x) \right]}{\textbf{E}_{p_t}\left[ \alpha_t(x)\,x \right]}.
\end{equation}
We found that, as for the Gini index, the coefficient of variation of the average evolution $\gamma_t(x)$ is a convex combination of the coefficients of variation of the two component functions $\alpha_t(x)\,x$ and $\beta_t(x)$.
Finally, we prove the lower bound:
\begin{align}
    \textbf{CV}_{\alpha x, t}^2 - \textbf{CV}_{x, t}^2 & = \frac{   \textbf{E}_{p_t}\left[ x \right]  }{   \textbf{E}_{p_t}\left[\alpha_t(x)\,x\right]^2   } \left[    \textbf{E}_{\tilde{p}_t}\left[\alpha_t(x)^2\,x \right]     -  \textbf{E}_{\tilde{p}_t}\left[x \right]  \textbf{E}_{\tilde{p}_t}\left[\alpha_t(x)\right]^2    \right] = \\
    & = \frac{   \textbf{E}_{p_t}\left[ x \right]  }{   \textbf{E}_{p_t}\left[\alpha_t(x)\,x\right]^2   } \left[ \textbf{Cov}_{\tilde{p}_t}\left[\alpha_t(x)^2,\, x \right] + \textbf{E}_{\tilde{p}_t}\left[x \right]\, \textbf{Var}_{\tilde{p}_t}\left[\alpha_t(x) \right] \right] \geq \\
    & \geq \frac{   \textbf{E}_{p_t}\left[ x^2 \right]  }{  \textbf{E}_{p_t}\left[ x \right]^2 }  \frac{   \textbf{Var}_{\tilde{p}_t}\left[\alpha_t(x) \right]   }{  \textbf{E}_{\tilde{p}_t}\left[\alpha_t(x)\right]^2} = \\
    & = \left( 1 + \textbf{CV}_{x,t}^2 \right)  \frac{   \textbf{Var}_{\tilde{p}_t}\left[\alpha_t(x) \right]   }{  \textbf{E}_{\tilde{p}_t}\left[\alpha_t(x)\right]^2} 
\end{align}
where we defined the new probability density:
\begin{equation}
    \tilde{p}_t(x)\coloneqq\frac{p_t(x)\,x}{\textbf{E}_{p_t}\left[ x \right]},
\end{equation}
and used the fact that a non-decreasing $\alpha_t(x)$ implies $\textbf{Cov}_{\tilde{p}_t}\left[\alpha_t(x)^2,\, x \right] \geq 0$.
Inserting the last result in equation \eqref{eq:convex_cv} and using the notation:
\begin{equation}
    \mathbf{\widetilde{CV}}_{\alpha, t}^2 \coloneqq  \frac{   \textbf{Var}_{\tilde{p}_t}\left[\alpha_t(x) \right]   }{  \textbf{E}_{\tilde{p}_t}\left[\alpha_t(x)\right]^2}, 
\end{equation}
we obtain:
\begin{equation}
    \textbf{CV}_{\gamma, t}^2 \geq  \frac{ 1 + \mathbf{\widetilde{CV}}_{\alpha, t}^2 }{ \left( 1 + \varrho_t \right)^2}\, \textbf{CV}_{x, t}^2 + \left( \frac{\varrho_t}{1 + \varrho_t} \right)^2 \textbf{CV}_{\beta, t}^2 + \frac{\mathbf{\widetilde{CV}}_{\alpha, t}^2}{ \left( 1 + \varrho_t \right)^2}.
\end{equation}
Hypothesis \ref{hyp:positive_dispersion} implies the existence of a value of $\kappa$ such that:
\begin{equation}
    x \geq \kappa \, \textbf{E}_{p_t}\left[  x \right] \Longrightarrow \textbf{Var}_{w_t}\left[ \left. x' \right| x \right] \geq \Gamma_t^2\, x^2.
\end{equation}
This implies that the noise term $\textbf{C}_{\Gamma, t}$ can be bounded as:
\begin{equation}
    \textbf{C}_{\Gamma, t}^2 \geq \frac{\Gamma_t^2}{g_t^2}  \, \left( 1 + \textbf{CV}_{x, t}^2\right) \, \textbf{P}_{t}\left( x \geq \kappa\, \textbf{E}_{p_{t}}\left[ x \right] \right)
\end{equation}
where we introduced the average growth factor:
\begin{equation}
g_t \coloneqq \frac{\textbf{E}_{p_t}\left[   \gamma_t(x) \right]}{\textbf{E}_{p_t}\left[  x \right]},
\end{equation}
and the probability to have wealth greater than $\kappa \, \textbf{E}_{p_t}\left[  x \right]$:
\begin{equation}
    \textbf{P}_{t}\left( x \geq \kappa\, \textbf{E}_{p_{t}}\left[ x \right] \right) \coloneqq \int_{\kappa\, \textbf{E}_{p_{t}}\left[ x \right]}^{\infty} dx' \, p_t(x')\, x'.
\end{equation}
Taking into account all these results in equation \eqref{eq:cv_evolution}, we obtain:
\begin{equation}
\label{eq:main_cv_bound}
    \textbf{CV}_{x, t+1}^2 \geq \left( \frac{ 1 + R_t   }{ 1 + \varrho_t } \right)^2 \textbf{CV}_{x, t}^2 + \left( \frac{\varrho_t}{1 + \varrho_t} \right)^2 \textbf{CV}_{\beta, t}^2 + \frac{\mathbf{\widetilde{CV}}_{\alpha, t}^2}{ \left( 1 + \varrho_t \right)^2} + \frac{\Gamma_t^2}{g_t^2}\,  \textbf{P}_{t}\left( x \geq \kappa\, \textbf{E}_{p_{t}}\left[ x \right] \right)
\end{equation}
with:
\begin{equation}
    \left(1 + R_t  \right)^2 = 1 + \mathbf{\widetilde{CV}}_{\alpha, t}^2 + \frac{\Gamma_t^2}{g_t^2}\,  \textbf{P}_{t}\left( x \geq \kappa\, \textbf{E}_{p_{t}}\left[ x \right] \right).
\end{equation}
To avoid an unbounded growth of wealth concentration, after some time $t^*$ the right hand side of this inequality has to be less than or equal to  $\textbf{CV}_{x, t}^2$ (or tend to it fast enough for $t \to \infty$).
While for the Gini index we found that a sufficient (and necessary) condition to avoid infinite concentration is $\varrho_t > 0$, here we find a stronger bound. 
The probability $\textbf{P}_{t}\left( x \geq \kappa\, \textbf{E}_{p_{t}}\left[ x \right] \right)$ can remain greater than zero for arbitrarily high values of $\textbf{CV}_{x, t}$, as can be seen for any distribution having finite average and diverging variance. 
Hence for the right hand side of inequality \eqref{eq:main_cv_bound} not to diverge, we must have at least:
\begin{equation}
    \left( 1 + \varrho_t \right)^2 >  1 + \mathbf{\widetilde{CV}}_{\alpha, t}^2 + \frac{\Gamma_t^2}{g_t^2}\,  \textbf{P}_{t}\left( x \geq \kappa\, \textbf{E}_{p_{t}}\left[ x \right] \right).
\end{equation}

\end{document}